\documentclass[journal]{IEEEtran}
\usepackage{amsmath,amssymb}

\usepackage{subcaption}
\usepackage{bm}
\usepackage{graphicx,graphics,color,psfrag}
\usepackage{cite,balance}
\usepackage{caption}
\allowdisplaybreaks
\usepackage{algorithm}
\usepackage{algorithmic}
\usepackage{accents}
\usepackage{amsthm}
\usepackage{url}
\usepackage[english]{babel}
\usepackage{multirow}
\usepackage{enumerate}
\usepackage{cases}
\usepackage{stfloats}
\usepackage{dsfont}
\usepackage{color,soul}
\usepackage{amsfonts}
\usepackage{cite,graphicx,amsmath,amssymb}
\usepackage{fancyhdr}
\usepackage{hhline}
\usepackage{graphicx,graphics}
\usepackage{array,color}
\usepackage{mathtools}
\usepackage{amsmath}
\usepackage{framed}
\usepackage{booktabs}  
\usepackage{multirow}  
\usepackage[most]{tcolorbox}
\newtcolorbox{mybeamdesign}[2][]{
	enhanced,
	colback=white,           
	colframe=black,          
	colbacktitle=white,      
	coltitle=black,          
	center title,
	 sharp corners,             
	boxrule=1pt,                 
	left=0pt, right=0pt,       
	top=0pt, bottom=0pt,         
	title={#2},                  
}

\newtheorem{definition}{\emph{\underline{Definition}}}

\newtheorem{lemma}{\emph{\underline{Lemma}}}

\newtheorem{proposition}{\emph{\underline{Proposition}}}

\newtheorem{example}{\emph{\underline{Example}}}
\newtheorem{remark}{\bf \emph{\underline{Remark}}}
\newtheorem{observation}{\emph{\underline{Observation}}}

\def\sinc{\mathrm{sinc}}

\def\l{\left}
\def\r{\right}
\def\({\left(}
\def\){\right)}

\def\b0{{\mathbf{0}}}

\graphicspath{{./Figs/}}
\begin{document}
	\captionsetup[figure]{name={Fig.}} 
	\title{Low-complexity Design for Beam Coverage\\
		in Near-field and Far-field: A Fourier \\
		Transform Approach}
	\author{Chao Zhou,
		Changsheng~You,
		Cong Zhou,
		Li Chen,
		Yi Gong,
		and Chengwen Xing
		\thanks{Chao Zhou, Changsheng~You, Cong Zhou, and Yi Gong are with the Department of Electronic and Electrical Engineering, Southern University of Science and Technology (SUSTech), Shenzhen 518055, China (e-mail: zhouchao2024@mail.sustech.edu.cn, youcs@sustech.edu.cn, zhoucong@stu.hit.edu.cn, and gongy@sustech.edu.cn).	
		Li Chen is with the Department of Electronic Engineering and Information Science, University of Science and Technology of China, Hefei 230026, China (e-mail: chenli87@ustc.edu.cn).
		Chengwen Xing is with the School of Information and Electronics, Beijing Institute of Technology, Beijing 100081, China (email: xingchengwen@gmail.com).
		\emph{(Corresponding author: Changsheng You.)}}\vspace{-20pt}
	} 
	
	\maketitle
	\begin{abstract}
		In this paper, we study efficient beam coverage design for multi-antenna systems in both far-field and near-field cases. To reduce the computational complexity of existing sampling-and-optimization based methods, we propose a new low-complexity yet efficient beam coverage design that exploits the Fourier-transform (FT) relationship between the antenna and spatial domains. To this end, we first formulate a general beam coverage optimization problem for both near-field and far-field cases to maximize the worst-case beamforming gain over a target region. For the far-field case, we show that the beam coverage design can be viewed as a spatial-frequency filtering problem, where angular coverage can be achieved by weight-shaping in the antenna domain via an inverse FT, yielding an infinite-length weighting sequence. Under the constraint of a finite number of antennas, a surrogate scheme is proposed by directly truncating this infinite-length sequence, which, however, inevitably introduces a \emph{roll-off effect} at the angular boundaries, yielding degraded worst-case beamforming gain. To address this issue, we characterize the finite-antenna-induced roll-off effect, based on which a \emph{roll-off-aware} design with a \emph{protective zoom} is developed to ensure a flat beamforming-gain profile within the target angular region. Next, we extend the proposed method to the near-field case. Specifically, by applying a first-order Taylor approximation to the phase of the near-field channel steering vector (CSV), the two-dimensional (2D) near-field beam coverage design (in both angle and inverse-range) can be transformed into a 2D inverse FT, leading to a low-complexity and closed-form beamforming design. Furthermore, an inherent near-field \emph{range defocusing effect} is observed, indicating that sufficiently wide angular coverage results in range-insensitive beam steering. Finally, numerical results demonstrate that the proposed FT-based approach achieves a comparable worst-case beamforming performance with that of conventional sampling-based optimization methods while significantly reducing the computational runtime by several orders-of-magnitude.
	\end{abstract}
	\vspace{-10pt}
	\section{Introduction}
	The evolution of multiple-input multiple-output (MIMO) technology represents a mainstream direction on the path toward next-generation wireless networks. For example, massive MIMO is a pivotal physical-layer enabler for fifth-generation (5G) that exploits enormous antenna resources to boost spectral efficiency, enhance coverage, and suppress inter-user/inter-cell interference~\cite{LuhqXLMIMO}. Likewise, extremely large-scale MIMO (XL-MIMO) is emerging  as a promising technology for sixth-generation (6G) that increases the number of antennas at base stations (BSs) by another order-of-magnitude, hence drastically enhancing the spectral efficiency and spatial resolution~\cite{YouNGAT}. Moreover, compared with massive MIMO that mostly serves users in the far-field with planar wavefronts, XL-MIMO significantly expands the Rayleigh distance, rendering the users more likely to be located in the near-field region featured by spherical wavefronts~\cite{zhang2022beam}. Such new channel characteristics enable an appealing function of \emph{beam-focusing} that focuses most beam energy around the user location/region. This property can be exploited to accommodate a variety of applications, such as high-precision localization~\cite{LuhqXLMIMO}, integrated sensing and communication~\cite{CongISAC}, and physical layer security~\cite{MyPLS}.
	Among others, beam coverage design plays a pivotal role in providing reliable connectivity across wide spatial regions, thereby supporting wireless sensing~\cite{LiRG}, beam tracking~\cite{Songtracking}, and hierarchical beam training~\cite{XiaozyCD,zhou2024near}.
	
	In this paper, we study beam coverage design in MIMO systems for both far-field and near-field cases. In view of existing works in this direction, they are mostly designed based on complex optimization methods and hence suffer from demanding computational complexity, or simply based on heuristic alternatives without performance guarantee. To address these issues,  we propose a new low-complexity beam coverage design based on a Fourier transform (FT) principle, which achieves comparable coverage performance  with the sampling-based optimization methods. 
	\vspace{-10pt}
	\subsection{Related Works}
	\subsubsection{Far-field case}
	In the far-field scenario, beam coverage design aims to achieve uniform beamforming gain over a specified angular region, which is often realized through beam broadening (or flat-top beam synthesis).
	Various methods have been proposed in the literature to achieve this goal, which can be categorized into optimization-based and heuristic methods. For optimization-based methods, the beam coverage design can be formulated as  optimization problems in different forms. For example, a worst-case beam coverage performance was formulated in~\cite{LiRG}, where a sampling-based optimization method was proposed to ensure continuous sensing coverage within a target region. Alternatively, another wide-beam design problem was considered in~\cite{Chen2020,Alkhateeb2014,Ning2023}, where the authors aimed to minimize the discrepancy between the generated beam pattern and the target reference pattern.
	While effective, these optimization-based methods generally require a massive number of angular samples to approximate continuous coverage and involve complicated matrix operations in optimization, hence incurring demanding computational complexity as the numbers of antennas and sampling points are sufficiently large. To reduce the computational complexity of optimization-based methods, several heuristic methods were developed. For example, one representative method is superimposing multiple discrete Fourier transform (DFT) codewords whose steering directions fall within the target angular interval for beam coverage design~\cite{Xu,Ning2023}. Another option is the chirp-sequence-based method~\cite{Peng2018}, which deliberately introduces a quadratic phase variation across the antenna array, thus broadening the  beamwidth in analog beamforming architectures.
	Although these heuristic methods provide a viable alternative, their performance may not be satisfactory in worst-case beam coverage scenarios.

	\subsubsection{Near-field case}
	In contrast to the far-field case, near-field beam coverage introduces several new challenges. First, unlike the far-field beam coverage in the angle domain only, the near-field counterpart requires coverage in both the angle and range domains~\cite{zhang2022beam}. Second, the huge number of antennas generally incur demanding computational complexity in the beam coverage design due to the involved matrix inversion and iterative optimization processes~\cite{LuhqXLMIMO}. Third, near-field spherical wavefronts feature quadratic phase variations in the near-field steering vector~\cite{Cui_CE}, rendering conventional far-field low-complexity beam coverage designs (e.g.,~\cite{Xu,Ning2023}) inefficient in the near-field case. To tackle these challenges, several optimization-based methods have been recently proposed. For instance, the authors in~\cite{Zheng2004} and~\cite{luo2025sphere} utilized the channel covariance matrix to optimize the beamforming vector for near-field beam coverage design. While effective in some scenarios, these methods primarily focus on improving average performance, and thus cannot ensure the worst-case performance. To further enhance beam coverage performance, the authors in~\cite{Wang2025} extended the sampling-based optimization method to the near-field scenario, by optimizing fully-digital beamformers to support hybrid beamforming architectures via iterative optimization.
	Following a similar optimization framework, a wavenumber-domain beam pattern design was proposed in~\cite{Weng26} to achieve satisfactory near-field beam coverage performance.
	To reduce the computational complexity of the above optimization-based methods~\cite{Wang2025,Weng26}, a heuristic beamforming coverage design was developed in~\cite{li2025Diverging}. By introducing additional quadratic phase variations into the beamforming vector, this method induces the so-called beam-diverging effect, effectively broadening the beamwidth with significantly reduced complexity.
	Despite these advancements, a comprehensive and systematic study of near-field beam coverage design is still lacking in the literature.

	\subsection{Motivations and Contributions}
	Motivated by the above, we study efficient beam coverage design of multi-antenna systems for both far-field and near-field cases, where a BS equipped with a uniform linear array (ULA) aims to provide beam coverage over a specified target region. Specifically, we formulate a general beam coverage optimization problem to maximize the worst-case beamforming gain over a prescribed target region. To reduce the computational complexity of existing optimization-based methods, we propose a new low-complexity yet efficient framework that exploits the FT principle to facilitate the beam coverage design in closed form for both near-field and far-field cases.
	The main contributions of this paper are summarized as follows.
	
	\begin{itemize}	
		\item First, for the far-field case, we show that the beam coverage design in the angular domain can be reformulated as a \emph{weight-shaping} problem in the antenna domain via  FT, which corresponds to an infinite-length antenna weighting sequence. Accounting for the finite number of antennas, a viable approach is truncating the antenna-domain sequence, leading to a surrogate scheme. However, such truncation inevitably induces a \emph{roll-off} effect at the angular boundaries, thereby degrading the beamforming gain. To address this issue, a \emph{roll-off-aware} beam coverage design is further proposed by introducing a \emph{protective zoom}  whose width scales inversely with the number of antennas, thereby preventing the roll-off behavior and ensuring a flat beamforming-gain profile within the target angular region.
			
		\item 
		Second, we extend the proposed FT-based method  to the near-field case, where beam coverage should be ensured jointly in the angle and range domains. Specifically, by applying a first-order Taylor expansion, the phase of near-field channel steering vector (CSV) is approximately linearized under moderate angular deviations. Under this approximation, the near-field beam coverage design over both angle and range domains can be reformulated as a \emph{weight-shaping} problem in the antenna domain via a two-dimensional (2D) inverse FT, thereby enabling a low-complexity beamforming design. Moreover, we reveal an inherent near-field \emph{range defocusing} effect, i.e., when the angular beam coverage becomes sufficiently wide, the near-field beam-focusing property degenerates into a range-insensitive beam steering, under which an angle-only beam coverage design is capable of providing broad range coverage.
		
		\item Finally, we discuss the extension of proposed method to more general scenarios, including multi-region beam coverage, near-field case with large angular deviations, and beam coverage based on analog beamforming architectures. Numerical results demonstrate that the proposed low-complexity beam coverage design achieves a comparable worst-case beamforming performance with those of high-complexity sampling-and-optimization based benchmarks, while significantly reducing the computational runtime by several orders-of-magnitude (e.g., from 885 ms to 0.0074 ms in the far-field case and from 1275 ms to 0.0292 ms in the near-field case), making it suitable for real-time applications such as beam training and tracking.
	\end{itemize}

	\emph{Notations:}  
	Bold lowercase letters (e.g., $\mathbf{a}$) denote vectors, and bold uppercase letters (e.g., $\mathbf{A}$) denote matrices. Calligraphic letters (e.g., $\mathcal{N}$) represent sets. The superscripts $(\cdot)^{H}$ and $(\cdot)^{T}$ denote the Hermitian (conjugate transpose) and transpose operations, respectively. The absolute value of a scalar is denoted by $|\cdot|$, and the Euclidean norm of a vector is denoted by $\|\cdot\|_{2}$. The symbol $\odot$ denotes the Hadamard product, while $\circledast$ represents the convolution operation. Finally, $\mathcal{O}(\cdot)$ denotes the standard big-$\mathcal{O}$ notation for characterizing algorithmic complexity.

	\section{System Model and Problem Formulation}\label{Sec2:label}
	We consider a multi-antenna communication system, 
	where a BS consisting of an $N$-antenna ULA (indexed by $\mathcal{N}\triangleq\{1,2,\ldots,N\}$) is deployed to provide beam coverage for a target region.\footnote{The proposed low-complexity beam coverage design can be extended to uniform planar array (UPA) systems. However, due to the coupling of horizontal and vertical angles in both near-field and far-field CSVs, obtaining the closed-form solution requires specific coordinate transformations to decouple the spatial angles, which is left for our future work.}
	The ULA is deployed along the $y$-axis with its center at the origin $(0,0)$ m. As such, the position of the $n$-th antenna is given by $\mathbf{u}_{n}= [0,u_{n}]^T,\forall n\in \mathcal{N}$, where $u_{n} = \frac{2n-N-1}{2}d$ with $d = \frac{\lambda}{2}$ denoting the half-wavelength inter-antenna spacing.
	
	\subsection{Channel Model}
	We consider two channel modeling cases depending on the array size. Specifically, when the number of antennas is relatively small, the far-field (planar-wavefront) channel model is adopted. However, when the array aperture becomes sufficiently large, the near-field (spherical wavefront) channel model needs to be considered to accurately characterize the electromagnetic propagation characteristics.
	\subsubsection{Far-field case}
	For this case, we consider the line-of-sight (LoS)-dominant channel model, which is widely used in high-frequency (e.g., mmWave and THz) bands, due to negligible power in the non-LoS (NLoS) paths~\cite{MacCartney2017ICC}.
	Accordingly, the far-field channel from the BS to an arbitrary user, denoted by $\mathbf{h}_{\rm FF}^{H}\in\mathbb{C}^{1\times N}$, can be modeled as
	\begin{align}\label{Exp:FFchannel}
		\mathbf{h}_{\rm FF}^H = \sqrt{N} \beta_{0} \mathbf{a}_{\rm FF}^H(\theta),
	\end{align}
	where $\beta_{0}$ and $\theta \in [-1,1]$ denote the complex-valued channel gain of the LoS path and its spatial angle, respectively. Additionally, the far-field CSV $\mathbf{a}_{\rm FF}(\theta)$ is given by
	\begin{align}\label{Exp:FFsteervec}
		\mathbf{a}_{\rm FF}(\theta)=\frac{1}{\sqrt{N}}
		\big[
		e^{\jmath\frac{2\pi}{\lambda}u_{1}\theta},
		\ldots,
		e^{\jmath\frac{2\pi}{\lambda}u_{N}\theta}\big]^T,
	\end{align}
	
	\subsubsection{Near-field case} When the array aperture is significantly large, the user is located in the near-field region of the BS, with its range $r$ satisfying $Z_{\rm Fres} \le r \le Z_{\rm Rayl}$, where $Z_{\rm Rayl}=\frac{2D^2}{\lambda}$ is the Rayleigh distance and $Z_{\rm Fres} = 0.5\frac{\sqrt{{D^3}}}{\lambda}$ is the Fresnel distance, and $D=(N-1)d$ represents the array aperture~\cite{LuhqXLMIMO}.
	For this case, the near-field LoS channel is modeled as follows based on spherical wavefronts 
	\begin{align}\label{Exp:NFchannel} 
		\mathbf{h}_{\rm NF}^H = \sqrt{N} \beta_{0} \mathbf{a}_{\rm NF}^H(\theta,r).
	\end{align} 
	Herein, the near-field CSV $\mathbf{a}_{\rm NF}(\theta,r)$ is given by
	\begin{align}
		\mathbf{a}_{\rm NF}(\theta,r)=\frac{1}{\sqrt{N}}\big[e^{-\jmath\frac{2\pi}{\lambda}(r_{1} - r)},\ldots,e^{-\jmath\frac{2\pi}{\lambda}(r_{N} - r)}\big]^T,
	\end{align}
	where $r_{n}$ is the distance between the user and the $n$-th BS antenna. Based on the Fresnel approximation~\cite{Cui_CE,Zhangtraing}, $r_{n}$ can be approximated as 
	\begin{align}
		r_{n} &= \sqrt{r^{2} + u_{n}^{2} - 2r u_{n} \theta} 
		\overset{(a)}{\approx} r - u_{n} \theta + \frac{u_{n}^{2}(1- \theta^{2})}{2 r} \nonumber \\
		& \triangleq \frac{1}{\xi} - u_{n} \theta  + \frac{u_{n}^{2}(1- \theta^2)}{2}\xi,
	\end{align}
	where $\xi \triangleq \frac{1}{r} \in \big[\frac{1}{Z_{\rm Rayl}},\frac{1}{Z_{\rm Fres}}\big]$ denotes the \emph{inverse} range of the user. Note that the inverse range $\xi$ is introduced to facilitate the subsequent low-complexity beam coverage design based on a linear approximation of near-field CSV with respect to (w.r.t.)~$\xi$ (see details in Section~\ref{Sec:NFcase}).
	Based on the above, the near-field CSV can also be expressed as a function of the spatial angle $\theta$ and inverse range $\xi$, i.e.,
	\begin{align}\label{Exp:NFCSV}
		\big[\mathbf{a}_{\rm NF}(\theta,\xi)\big]_{n}=\frac{1}{\sqrt{N}}e^{\jmath\frac{2\pi}{\lambda}\big(u_{n} \theta - \frac{u_{n}^{2}(1-\theta^{2})}{2}\xi \big)}, \forall n\in \mathcal{N}.
	\end{align}

	\subsection{Problem Formulation}
	In this work, we aim to maximize the minimum beamforming gain (or equivalently power gain) across a specified target region, which has been widely considered in the existing literature (see, e.g.,~\cite{Ning2023,Chen2020,Wang2025}) for ensuring reliable beam coverage (with detailed applications discussed in Section~\ref{Sec:dis}).
	Specifically, for the far-field case, the target region is defined~as
	\begin{align}\label{Exp:FF_TR}
		\mathcal{A}_{\rm FF}\triangleq \big\{\theta \mid\theta \in \Theta \big\}, 
	\end{align}
	where $\Theta \triangleq [\theta_{\min}, \theta_{\max}]$ denotes the considered spatial angle interval, with $\theta_{\min}$ and $\theta_{\max}$ representing the minimum and maximum target spatial angles, respectively. Let $\mathbf{w}=[w_{1},\ldots,w_{N}]^T$ denote the fully-digital transmit beamforming vector of the BS.\footnote{We consider the fully-digital beamforming architecture in this work, while the extension to a purely analog beamforming architecture (e.g., phased array) will be discussed in Section~\ref{Sec:dis}.}  Then, the beamforming gain for the far-field case is given by
	$	g_{\rm FF}(\theta;\mathbf{w})  \triangleq 
		| \mathbf{a}_{\rm FF}^H(\theta) \mathbf{w}|$.
	For the near-field case, the beam coverage should be considered over both the angle and range domains. As a result, the target region becomes a 2D region characterized as 
	\begin{align}\label{Exp:NF_TR}
		\mathcal{A}_{\rm NF} \triangleq \big\{(\theta,\xi) \mid \theta \in \Theta, \xi \in \Xi \big\}, 
	\end{align}
	where $\Xi \triangleq [\xi_{\min}, \xi_{\max}]$ denotes the inverse-range interval, with $\xi_{\min}$ and $\xi_{\max}$ representing the minimum and maximum target inverse ranges, respectively. Similarly, the beamforming gain for the near-field case is defined as
	$	g_{\rm NF}\big((\theta,\xi);\mathbf{w}\big) \triangleq
		| \mathbf{a}_{\rm NF}^H(\theta,\xi) \mathbf{w}|$. 
	
	Based on the above, to ensure reliable beam coverage over a specified target region (i.e., $\mathcal{A}_{\chi}, \chi \in \{\rm{FF}, \rm{NF}\}$),  the optimization problem to maximize the \emph{worst-case} beamforming gain is formulated as follows
	\begin{align}
		\textbf{(P1)}:\;\max_{\mathbf{w}}& \;\; \min_{\mathcal{U}_{\chi} \in \mathcal{A}_{\chi}} \;\; g_{\chi}\big(\mathcal{U}_{\chi};\mathbf{w}\big)  \nonumber \\
		{\rm {s.t.}}& \;\; \quad \|\mathbf{w}\|_{2}^{2} \le P_{\rm t}, \label{C:Power} 
	\end{align}
	where $P_{\rm t}$ represents the maximum BS transmit power, and $\chi = \rm{NF}$ and $\chi = \rm {FF}$ correspond to the near-field and far-field cases, respectively. Additionally,  $\mathcal{U}_{\rm FF}\triangleq \{\theta\}$ and $\mathcal{U}_{\rm NF}\triangleq \{(\theta,\xi)\}$ represent the corresponding spatial domains for each case.
	Note that Problem \textbf{(P1)} is a non-convex optimization problem which is difficult to optimally solve in general, since 1) the objective function is non-convex w.r.t. the beamforming vector $\mathbf{w}$, and 2) the worst-case beamforming gain in the objective function involves infinitely many candidate locations in the continuous uncertainty set $\mathcal{A}_{\chi}$. 
	
	To tackle the above difficulties, we first reformulate it into the following equivalent form
	\begin{align}
		(\textbf{P2}):\;\max_{\mathbf{w}}&\quad \gamma_{\chi} \nonumber \\
		{\rm {s.t.}}&\quad \eqref{C:Power}, \nonumber   \\
		&\quad g_{\chi}\big(\mathcal{U}_{\chi};\mathbf{w}\big)  \ge \gamma_{\chi},\;\forall\;\mathcal{U}_{\chi} \in \mathcal{A}_{\chi} ,  \label{C:Gain}
	\end{align} 
	where $\gamma_{\chi}, \chi\in \{\rm FF,\rm NF\}$ is an introduced auxiliary variable. To deal with the non-convex constraint in~\eqref{C:Gain}, existing works (e.g.,~\cite{Ning2023,Chen2020}) mostly re-expressed it as a convex constraint by discretizing the target region and applying successive convex approximation (SCA) techniques.
	However, this method inevitably incurs prohibitively high computational complexity in practice, which increases \emph{linearly} with both the numbers of antennas and sampling points. 
	To address this issue, in the following sections, we first consider the far-field case for which a low-complexity yet efficient method is proposed to maximize the worst-case beamforming gain in the target spatial region. Then, we further extend it to the near-field case by accounting for the additional range domain.
	\vspace{-6pt}
	\section{Far-field Beam Coverage}\label{Sec:FFcase}
	In this section, we propose a new and efficient FT-based far-field beam coverage design, which enjoys a significantly lower computational complexity than existing optimization-based methods.
	Specifically, we show that the far-field beam coverage problem in the angle domain can be equivalently re-expressed as a \emph{spatial–frequency filtering} problem, where the antenna-domain beamforming and the far-field beam coverage are linked via FT. Moreover, to tackle the \emph{roll-off} effect that arises from the finite number of antennas and results in beamforming gain loss at the angular boundary of target region, a protective zoom is deliberately designed to prevent roll-off inside the desired target region.
	
	\vspace{-6pt}
	\subsection{FT between Spatial-frequency and Antenna Domains}
	To facilitate the low-complexity beam coverage design, we first study the intrinsic relationship between the beamforming vector in the antenna domain and the resulting beam pattern in the angle domain. Specifically, by characterizing the far-field CSV over the target region $\mathcal{A}_{\rm FF}$, we show that there is an FT relationship between the spatial-frequency and the antenna domains.

	\subsubsection{Far-field CSV}\label{Sec:FF-B1}
	To tackle the non-convex constraint~\eqref{C:Gain}, we first characterize the far-field CSV within $\mathcal{A}_{\rm FF}$ as follows.
	\begin{lemma}[Far-field CSV]\label{Lem:FF_CSV} \rm 
		Given the target region $\mathcal{A}_{\rm FF}$ and a reference angle $\theta_{0} \in \mathcal{A}_{\rm FF}$, the far-field CSV w.r.t. an arbitrary angle $\theta \in \mathcal{A}_{\rm FF}$ can be expressed as
		\begin{align}
			\big[\mathbf{a}_{\rm FF}(\theta)\big]_{n} = \frac{1}{\sqrt{N}} \exp\Big(\jmath\frac{2\pi}{\lambda}u_{n}(\theta_{0} + \Delta \theta)\Big),\forall n\in\mathcal{N},
		\end{align}
		where $\theta = \theta_{0} + \Delta \theta$.
	\end{lemma}
	
	\subsubsection{Normalized beamforming gain}
	Based on {\bf Lemma~\ref{Lem:FF_CSV}}, the far-field CSV of any angle $\theta \in \mathcal{A}_{\rm FF}$ can be expressed as a function of the reference angle $\theta_{0}$ and the angle difference $\Delta \theta$. As such, the beamforming gain for an arbitrary spatial angle can be re-expressed as follows.
	\begin{lemma}[FT between antenna domain and spatial frequency domain]\label{Lem:FTrel}\rm
		For the far-field case, given a transmit beamforming vector $\mathbf{w}_{\rm FF} $, the (normalized) beamforming gain at~$\theta = \theta_{0} + \Delta \theta$ (i.e., $|\mathbf{a}_{\rm FF}^{H}(\theta)\mathbf{w}_{\rm FF}|$) 
		is given by
		\begin{align}
			&g_{\rm FF}(\theta) =  |\mathbf{a}_{\rm FF}^{H}(\theta)\mathbf{w}_{\rm FF}| = \Big|\frac{1}{\sqrt{N}}\sum_{n=1}^{N}  w_{{\rm FF},n} e^{-\jmath\frac{2\pi}{\lambda}u_{n}(\theta_{0} + \Delta \theta)}\Big| \nonumber \\
			&\quad = \Big|\frac{1}{\sqrt{N}}\sum_{n=1}^{N}  v_{{\rm FF},n}  e^{-\jmath\frac{2\pi}{\lambda}u_{n} \Delta \theta} \Big| \triangleq |f_{\rm FF}(\Delta \theta;\theta_{0},N)|,\label{Exp:FF_Fourier}
		\end{align}
		where  $v_{{\rm FF},n} \triangleq w_{{\rm FF},n}e^{-\jmath \frac{2\pi}{\lambda}u_{n} \theta_{0} }, \forall n\in \mathcal{N}$, or equivalently, we have $\mathbf{w}_{\rm FF} = \mathbf{a}_{\rm FF}(\theta_{0}) \odot \mathbf{v}_{\rm FF}$ with $\mathbf{v}_{\rm FF} = [v_{{\rm FF},1}, \ldots, v_{{\rm FF},N}]^{T}$.
	\end{lemma}
	One can observe from~\eqref{Exp:FF_Fourier} that $ f_{\rm FF}(\Delta \theta;\theta_{0})$ is essentially a DFT of $\{v_{{\rm FF},n}\}_{n=1}^{N}$. This equivalence establishes an FT relationship, where the antenna positions $u_{n}$ are treated as discrete time samples, the vector $\mathbf{v}_{\rm FF}$ serves as the time-domain signal, and the spatial angle deviation $\Delta \theta$ resembles the frequency variable. This reveals an inherent FT relationship between the antenna domain and spatial-frequency domain, implying that the beam coverage design in the angular domain can be effectively achieved by a corresponding spectral filterer design in the antenna domain.
	
	\vspace{-14pt}
	\subsection{Proposed Surrogate Far-field Beam Coverage Design}  
	Based on the above insight, we first propose a surrogate far-field beam coverage design in this subsection. Then, we show that the finite number of antennas result in an inevitable \emph{roll-off effect}, and hence a degraded beamforming gain at the angular boundary of the target region.
	\subsubsection{Problem reformulation}\label{Sec:FF2} Given {\bf Lemma~\ref{Lem:FTrel}}, we first reformulate the original Problem \textbf{(P2)} by considering an ideal rectangular beamforming gain profile over the target region.
	
	Specifically, without loss of generality, we select the geometric center of
	the target angular region as the reference angle, i.e.,
	$	\theta_{0} =\big({\theta_{\min} + \theta_{\max}}\big)/{2}$, 
	such that the angle deviation in the target region satisfies
	\begin{align}
		|\Delta \theta|\le {\big(\theta_{\max} - \theta_{\min}\big)}/{2} \triangleq \mu. 
	\end{align}
	As such, based on~\eqref{Exp:FF_Fourier}, constraint~\eqref{C:Gain} under the far-field case can be rewritten as
	\begin{align}\label{Exp:FF_C}
		|f_{\rm FF}(\Delta \theta;\theta_{0},N) | \ge \gamma_{\rm FF},~\forall\;
		|\Delta \theta| \le \mu. 
	\end{align}
	To satisfy constraint~\eqref{C:Gain} and ensure the worst-case beamforming gain performance, we construct a beam pattern with an ideally flat and rectangular profile over the target region. Moreover, to facilitate the low-complexity beam coverage design, we enforce that ${f}_{\rm FF}$ is real-valued. As such, a target beamforming gain in the considered spatial region is given by 
	\begin{equation}\label{Exp:FF_window}
		{f}_{\rm FF}(\Delta \theta;\theta_{0},N) = \left\{
		\begin{aligned}
			&{\gamma_{\rm FF}}, && \forall\;|\Delta \theta| \le \mu,\\
			& 0,     &&\textrm{otherwise}.\\
		\end{aligned}
		\right. 
	\end{equation}
	Note that although the real-valued condition limits the DoFs in the beamforming design, it shall be shown in Section~\ref{Sec:NR} to achieve comparable coverage performance with the sampling-based method.
	
	Based on the above, the beam coverage design over the target region $\mathcal{A}_{\rm FF}$ can be reformulated as  
	\begin{align}
		(\textbf{P3}):\;\max_{\mathbf{w}_{\rm FF}}~\gamma_{\rm FF} \quad
		{\rm {s.t.}}~ \eqref{C:Power},~\eqref{Exp:FF_window}\nonumber.
	\end{align} 
	
	\subsubsection{Proposed surrogate scheme} Note that Problem \textbf{(P3)} can be viewed as designing a \emph{rectangular window} in the spatial angle domain, thus the vector $\mathbf{v}_{\rm FF}$ can be efficiently obtained by a two-step surrogate method elaborated below.
	
	{\bf Step 1} (Inverse FT): Given the \emph{rectangular window-like} beamforming gain over the target region $\mathcal{A}_{\rm FF}$, the ideal antenna-domain sequence $[\bar{\mathbf{s}}_{{\rm FF}}]_{n}, n\in \mathcal{Z}$ (assuming an infinite number of antennas) that is able to achieve the desired flat beam coverage over $\mathcal{A}_{\rm FF}$ in~\eqref{Exp:FF_window} can be obtained via the following inverse FT 
	\begin{align}\label{Exp:FF_weightvec}
		[\bar{\mathbf{s}}_{\rm FF}]_{n} \triangleq \bar{s}_{{\rm FF}, n} = 2 \mu {\gamma_{\rm FF}} \sinc\big(\frac{2 u_{n}\mu}{\lambda}\big), \forall n\in \mathcal{Z},
	\end{align}
	where $\sinc(a)= \frac{\sin(\pi a)}{\pi a}$. The proof of~\eqref{Exp:FF_weightvec} is established in Appendix~\ref{App:IdealWS}. 
	
	{\bf Step 2} (Antenna domain truncation): Let $[\bar{\mathbf{v}}_{{\rm FF}}]_{n} \triangleq \bar{v}_{{\rm FF}, n}  = \sinc\big(\frac{2 u_{n}\mu}{\lambda}\big), n\in \mathcal{Z}$.
	Then, the $N$-dimensional vector $\mathbf{v}_{\rm FF}$ can be obtained by truncating the ideal infinite sequence $\bar{\mathbf{v}}_{{\rm FF}}$ for the case of an infinite number of antennas with the array mask $\mathbf{m}$. Mathematically, the vector $\mathbf{v}_{\rm FF}$ is given by
	\begin{align}\label{Exp:FFWSvec}
		[\mathbf{v}_{\rm FF}]_{n} = \alpha \cdot [\bar{\mathbf{v}}_{{\rm FF}}\odot \mathbf{m}]_{n}= \alpha\cdot  \sinc\big(\frac{2 u_{n}\mu}{\lambda}\big), n\in\mathcal{N},
	\end{align}
 	where $\alpha = \sqrt{ \frac{P_{\rm t}}{ \sum_{n\in\mathcal{N}} \left| \sinc\big(\frac{2 u_{n}\mu}{\lambda}\big) \right|^2 } }$ is an auxiliary variable for satisfying the transmit power constraint and $\mathbf{m}$ denotes a rectangular window mask whose elements are given by
    \begin{equation}\label{Exp:Win}
    	[\mathbf{m}]_{n} \triangleq	m_{n} =
    	\begin{cases}
    		1, & \text{if } n \in \mathcal{N}, \\
    		0, & \text{if } n\in\mathcal{Z}~\text{and}~n\notin \mathcal{N}.
    	\end{cases}
    \end{equation}
 
	Note that in~\eqref{Exp:FFWSvec}, the vector $\mathbf{v}_{\rm FF}$ functions as an \emph{amplitude-shaping} vector specifying the power distribution across antennas to synthesize the rectangular beamforming gain centered at $\theta = 0$. 
	Based on the relationship $\mathbf{w}_{\rm FF} = \mathbf{a}_{\rm FF}(\theta_{0}) \odot \mathbf{v}_{\rm FF}$, the resultant beamforming vector of the proposed surrogate method is obtained as follows.
	\begin{mybeamdesign}{\textbf{Surrogate beam coverage design}}
		{\bf Step 1} (Weight-shaping design):
		\begin{align}
			[\mathbf{v}_{\rm FF}]_{n} =  \alpha\cdot  \sinc\big(\frac{2 u_{n}\mu}{\lambda}\big), n\in\mathcal{N}.
		\end{align}
		{\bf Step 2} (Beamforming design):
		\begin{align}
		\mathbf{w}_{\rm FF} = \mathbf{a}_{\rm FF}(\theta_{0}) \odot \mathbf{v}_{\rm FF}. 
		\end{align}
	\end{mybeamdesign}

	\begin{example} \rm 
		\begin{figure}[t]
			\centering
			\begin{subfigure}[b]{0.49\linewidth}
				\includegraphics[width=1\linewidth]{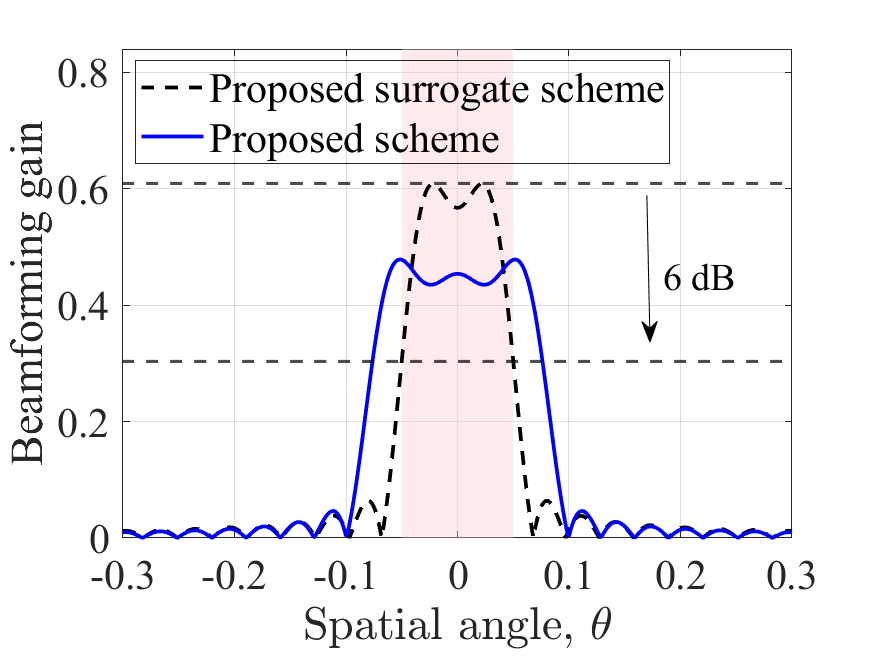}
				\caption{Beamforming gain.}
				\label{Fig:FFcasePattern}
			\end{subfigure}
			\hspace{-5pt}
			\begin{subfigure}[b]{0.46\linewidth}
				\includegraphics[width=1\linewidth]{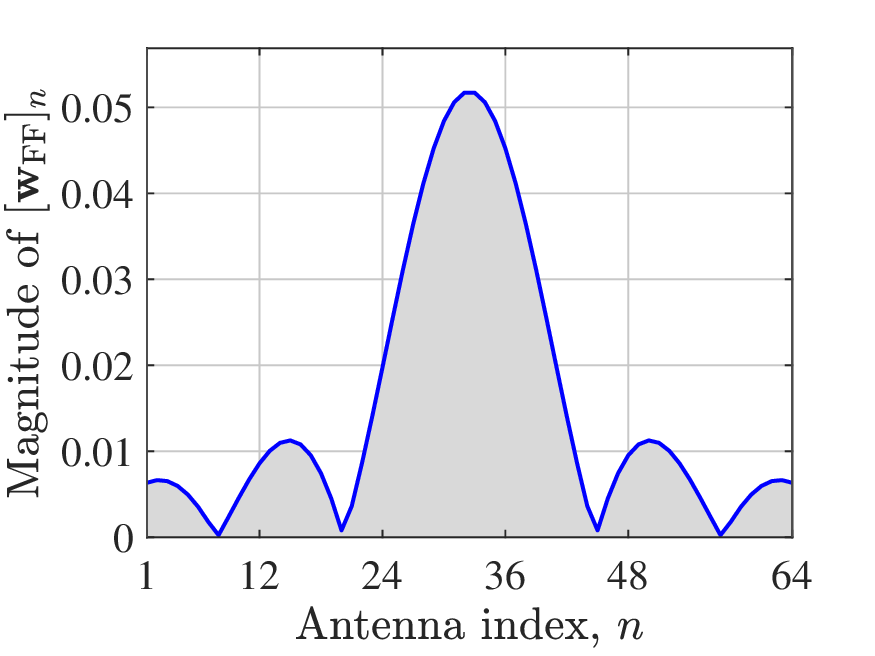}
				\caption{Magnitude of $[\mathbf{w}_{\rm FF}]_{n}$.}
				\label{Fig:FFcasePower}
			\end{subfigure}
			\caption{Beamforming gain and magnitude of $[\mathbf{w}_{\rm FF}]_{n}$.}
			\label{Fig:FFcase}
			\vspace{-14pt} 
		\end{figure}
		
		In Fig.~\ref{Fig:FFcase}, we numerically present the beamforming gain and power allocation to different antennas for the far-field case given $\theta_{\min} =-0.05$,  $\theta_{\max} =0.05$, and $N=64$. Several key observations are summarized below:
		\begin{itemize}
			\item \emph{Rectangular gain profile}: 
			The proposed surrogate scheme can be used to achieve a flat and rectangular-window-like beamforming gain that covers the target angular region.
			
			\item \emph{$\sinc(\cdot)$-shaped power distribution}: 
			The magnitude of the designed beamforming vector across the antenna domain exhibits a $\sinc(\cdot)$-like shape, as presented in~\eqref{Exp:FF_weightvec}.
			
			\item \emph{Roll-off property}: 
			Due to the finite number of antennas, the main-lobe of beamforming gain pattern generally exhibits a \emph{non-ideal roll-off}, hence inevitably causing a beamforming gain loss (about $6$ dB) at the angular boundaries of the target region. 
		\end{itemize} 
	\end{example}

	\subsection{Proposed Beam Coverage Design for Far-field Case}\label{Sec:ProFFscheme}
	In this subsection, we first analytically characterize the roll-off effect of the proposed surrogate method, based on which a \emph{roll-off-aware} beam coverage design method is devised by judiciously enlarging the target spatial region via a \emph{protective zoom}, thus improving the beamforming gain at the spatial boundary of the original target region (see Fig.~\ref{Fig:FFcase}(a)).
	
	Specifically, for ULAs with a finite number of antennas, the weight-shaping vector $\mathbf{v}_{\rm FF}$ in~\eqref{Exp:FFWSvec} can be interpreted as applying a rectangular window function $\mathbf{m}$ to the ideal weight sequence $\bar{\mathbf{v}}_{\rm FF}$, which can be equivalently rewritten as
	\begin{align}
		{v}_{{\rm FF},n}^{} = \bar{v}_{{\rm FF},n}^{} \cdot m_{n},
	\end{align}
	with $m_{n}$ defined in~\eqref{Exp:Win}. In both far-field and near-field cases (see Section~\ref{Sec:NFcase}), the weight-shaping vectors $\mathbf{v}_{\chi},\chi\in \{\rm FF,\rm NF\}$ reconfigure the power allocation across antennas, where the parameter $\alpha$ is introduced to satisfy the transmit power constraint. Since the value of $\alpha$ does not affect the beam coverage region, the beamforming gain achieved without considering the transmit power constraint (i.e., setting $\alpha = 1$) is referred to as the \emph{power-unconstrained beamforming gain} (shortened as unconstrained beamforming gain), which characterizes the beam coverage pattern generated by the beamforming vector of the surrogate method. As such, the rectangular window function $\mathbf{m}$ (i.e., a finite number of antennas) leads to the following unconstrained beamforming gain.
	\begin{lemma}[Unconstrained beamforming gain]\label{Lem:Uncon_gain}\rm Given the weight-shaping vector $\mathbf{v}_{\rm FF}$ in~\eqref{Exp:FFWSvec}, the unconstrained beamforming gain within the desired beam coverage region can be written as follows in a convolution form
		\begin{align}\label{Exp:Conv}
			\left| {f}_{\rm FF}^{(\rm unc)}(\Delta \theta,N)\right|  &\triangleq \left| \mathbf{a}_{\rm FF}^{H}(\theta_0 + \Delta \theta)\cdot \big(\mathbf{a}_{\rm FF}(\theta_{0}) \odot \mathbf{v}_{\rm FF}\big)\right|  \nonumber\\
			& =\frac{1}{N} \left|   f_{\rm \bar{v}}(\Delta \theta)  \circledast   \Omega(\Delta \theta,N) \right| \nonumber \\
			& = \frac{1}{2 \mu N} \int_{-\mu}^{\mu} N \frac{\sin\big(\frac{N \pi (\Delta\theta - x)}{2}\big)}{\frac{N \pi (\Delta\theta - x)}{2}} \, dx,
		\end{align}
		where $f_{\rm \bar{v}}(\Delta \theta) = 
		\begin{cases} 
			\frac{1}{2\mu}, & \text{if } |\Delta \theta| \leq \mu, \\
			0, & \text{if } |\Delta \theta| > \mu,
		\end{cases}$ is the FT of $\bar{\mathbf{v}}_{\rm FF}$ and  $\Omega(\Delta \theta,N) \!\triangleq\! \sum_{n=1}^{N}  e^{-\jmath\frac{2\pi}{\lambda}u_{n} \Delta \theta} =e^{-\jmath\pi\Delta \theta(N+1)/2} \frac{\sin(N\pi\Delta \theta/2)}{\sin(\pi\Delta \theta/2)} $ is the Dirichlet kernel induced by $\mathbf{m}$.
	\end{lemma}
	\begin{proof}
		Please refer to Appendix~\ref{App:UAG}
	\end{proof}
	
	To better understand the roll-off behavior of the unconstrained beamforming gain near the spatial boundary, we next present an approximation for the unconstrained beamforming gain that characterizes the roll-off property according to the value of $\mu$.
	
	\begin{proposition}[Roll-off property]\label{Pro:Rolloff}\rm
		Given the maximum angle deviation $\mu > \frac{2}{N}$,
		the unconstrained beamforming gain in \eqref{Exp:Conv} can be further approximated as\footnote{Given that practical beam coverage requirements typically exceed the angular resolution (i.e., $2/N$), we consider the scenario in which the maximum angular deviation satisfies $\mu > 2/N$.}  
		\begin{align}
			\big|{f}_{\rm FF}^{(\rm unc)}(\Delta \theta,N)\big| \approx \frac{1}{2 \mu N} \times
			\begin{cases} 
				2,\quad\quad\quad\quad\;\text{if }|\Delta \theta| \leq \mu-\frac{2}{N}, \\[5pt]
				1-\frac{2}{\pi}\text{Si}\big(\frac{N\pi}{2}(|\Delta\theta|-\mu)\big), \\
				\;\;\;\;\text{if }\mu-\frac{2}{N}<|\Delta \theta|<\mu+\frac{2}{N}, \\[5pt]
				0,\quad\quad\quad\quad\;\text{if }|\Delta \theta| \geq \mu+\frac{2}{N}, \\
			\end{cases} \nonumber
		\end{align}
		where $\text{Si}(a)\triangleq \int_{0}^{a}\frac{\sin t}{t}\,dt$. This approximation exploits the asymptotic behavior that $\text{Si}(a)\approx \pm \frac{\pi}{2}$ when $|a| \ge \pi$.
	\end{proposition}
	\begin{proof}
		Please refer to Appendix~\ref{App:Sifun}.
	\end{proof}
	\begin{example}\rm
		\begin{figure}[t]
			\centering
			\includegraphics[width=0.35\textwidth]{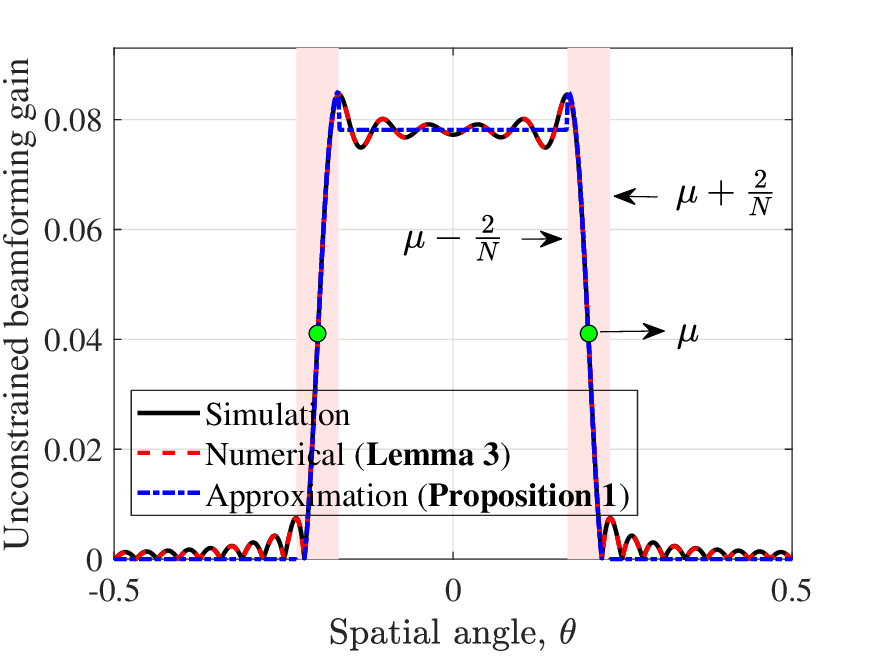}
			\caption{Unconstrained beamforming gain in the angle domain given $\theta_{\min} =-0.2$,  $\theta_{\max} =0.2$, $f = 30$ GHz, and $N=64$.} \label{Fig:Rolloff}
			\vspace{-14pt}
		\end{figure}
		In Fig.~\ref{Fig:Rolloff}, we present the unconstrained beamforming gain and power distribution for the far-field case given $\theta_{\min} =-0.2$,  $\theta_{\max} =0.2$, $f = 30$ GHz, and $N=64$. It is observed that the unconstrained beamforming gain, as given in {\bf Lemma~\ref{Lem:Uncon_gain}}, is consistent with the result obtained using the FT-based beam coverage design. Moreover, by approximating the unconstrained beamforming gain with the piecewise function in {\bf Proposition~\ref{Pro:Rolloff}}, we thus identify the roll-off interval, which is highlighted in pink (see Fig.~\ref{Fig:Rolloff}).
	\end{example}
	
	{\bf Proposition~\ref{Pro:Rolloff}} means that, given the maximum angle deviation $\mu > \frac{2}{N}$, the roll-off behavior occurs when $\mu-\frac{2}{N}<|\Delta \theta|<\mu+\frac{2}{N}$. Note that the width of this roll-off region is $\frac{4}{N}$, which is inversely proportional to the number of antennas $N$. 
	Moreover, when $|\Delta \theta| = \mu$, the unconstrained beamforming gain reduces to	$|{f}_{\rm FF}^{(\rm unc)}(\Delta \theta)| = 1$  due to $\text{Si}\big(0\big)=0$. This corresponds to a $-6$ dB reduction compared to the value when $|\Delta \theta| \leq \mu-\frac{2}{N}$.
	
	To prevent roll-off inside the target region, we propose to enlarge the design region by adding a protective zoom of width $2/N$, leading to the following beamforming design. Specifically, given the target region $\mathcal{A}_{\rm FF}$ and the reference angle~$\theta_{0}$, the proposed \emph{roll-off-aware} beam coverage design, obtained by solving Problem \textbf{(P3)}, follows a three-step process as detailed below.
	\begin{mybeamdesign}{\textbf{\emph{Roll-off-aware} beam coverage design}}
	{\bf Step 1} (Add protective zoom): 
	\begin{align}
		\mu^{+} = \mu + {2}/{N}. 
	\end{align}	
 	{\bf Step 2} (Weight-shaping design): 
 	\begin{align}
 		 [\tilde{\mathbf{v}}_{\rm FF}]_{n} = \tilde{{v}}_{{\rm FF},n} = \alpha \cdot \sinc\big(\frac{2 u_{n}\mu^{+}}{\lambda}\big), \forall n\in\mathcal{N}.
 	\end{align}
  	{\bf Step 3} (Beamforming design): 
	\begin{align}
		\tilde{\mathbf{w}}_{\rm FF} =  \mathbf{a}_{\rm FF}(\theta_{0}) \odot \tilde{\mathbf{v}}_{\rm FF}.
	\end{align}
\end{mybeamdesign}
The introduction of protective zoom ensures that the roll-off behavior occurs outside the target region, which in turn guarantees the worst-case beamforming gain over $\mathcal{A}_{\rm FF}$.

	\begin{remark}[Effect of number of antennas]\rm
		{\bf Proposition~\ref{Pro:Rolloff}} indicates that a protective zoom of width $\frac{2}{N}$ is necessary to prevent roll-off inside the target region, which is inversely proportional to the number of antennas. Notably, as $N$ approaches infinity, the width of the protective zoom converges to zero, which is expected since an array with an infinite aperture can achieve a perfect rectangular beam profile without any roll-off, rendering the protective zoom unnecessary in this ideal case.
	\end{remark}
	\vspace{-10pt}
	\section{Near-field Beam Coverage}\label{Sec:NFcase}
	In this section, we further consider the near-field beam coverage design. Compared to the far-field case, the more complex near-field CSV and the requirement of 2D (angle and range) coverage make it more challenging to solve Problem \textbf{(P2)}. To tackle these challenges, we extend the proposed far-field beam coverage design to the near-field case by leveraging near-field CSV approximation and 2D inversion FT. Additionally, we show an interesting result that when the angular beam coverage is sufficiently large, the near-field beam cannot focus energy well in the range domain, leading to a new wide-beam induced \emph{near-field defocusing effect}.
	\vspace{-10pt}
	\subsection{Proposed Beam Coverage Design for Near-field Case}
	Similar to the far-field case, we first approximate the near-field CSV based on a first-order Taylor expansion,\footnote{Although the second-order expansion provides higher accuracy, the resulting quadratic phase terms  break the FT relationship between the antenna and spatial-frequency domains, thereby complicating low-complexity beamforming design. In contrast, the proposed first-order expansion preserves this FT relationship while still capturing the range-dependent phase variations.} and then develop an efficient FT-based near-field beam coverage design.
	
	\subsubsection{Approximated near-field CSV and beamforming gain} 
	For this case, the near-field CSV in the target 2D region (see~\eqref{Exp:NFCSV}) w.r.t. $\theta$ and $\xi$ can be approximated as follows.
	\begin{lemma}[Approximated near-field CSV]\label{Lem:approx_CSV}
		\rm 
		For the target region $\mathcal{A}_{\rm NF}$ and by denoting $(\theta_{0},\xi_{0})\in \mathcal{A}_{\rm NF}$ a reference point, the CSV corresponding to any point  $(\theta,\xi)\in \mathcal{A}_{\rm NF}$ can be approximated as
		\begin{align}\label{Exp:approx_CSV}
			\big[\mathbf{a}_{\rm NF}(\theta,\xi)\big]_{n} &\approx \frac{1}{\sqrt{N}} \exp\Big({\jmath \frac{2\pi}{\lambda}\phi_{n}(\theta_{0},\xi_{0},\Delta\theta,\Delta \xi)}\Big)\nonumber \\
			&\triangleq \big[\mathbf{a}_{\rm NF}^{(\rm app)}(\theta_0 +\Delta \theta,\xi_0 + \Delta \xi )\big]_{n},\forall n\in\mathcal{N}
		\end{align}
		where $\Delta\theta=\theta-\theta_{0}$,  $\Delta\xi=\xi-\xi_{0}$, and
		\begin{align}
			&\phi_{n}(\theta_{0},\xi_{0},\Delta\theta,\Delta \xi) 
			=\underbrace{ u_{n}\theta_{0} - \frac{u_{n}^{2}(1-\theta_{0}^{2})}{2}\xi_{0} }_{\varphi_{n}(\theta_{0},\xi_{0})} + \nonumber \\
			&  
			\underbrace{\Big( u_{n}+u_{n}^{2}\theta_{0} \xi_{0} \Big)}_{\zeta_{n}^{(\theta)}(\theta_{0},\xi_{0})}\Delta\theta + \underbrace{ \Big( -\frac{u_{n}^{2}(1-\theta_{0}^{2})}{2} \Big) }_{\zeta_{n}^{(\xi)}(\theta_{0})}\Delta \xi, \forall n\in \mathcal{N}.
		\end{align}
	\end{lemma}
	\begin{proof}
		Please refer to Appendix~\ref{App:1}.
	\end{proof}
	
\begin{figure}[t]
		\centering
		\includegraphics[width=0.35\textwidth]{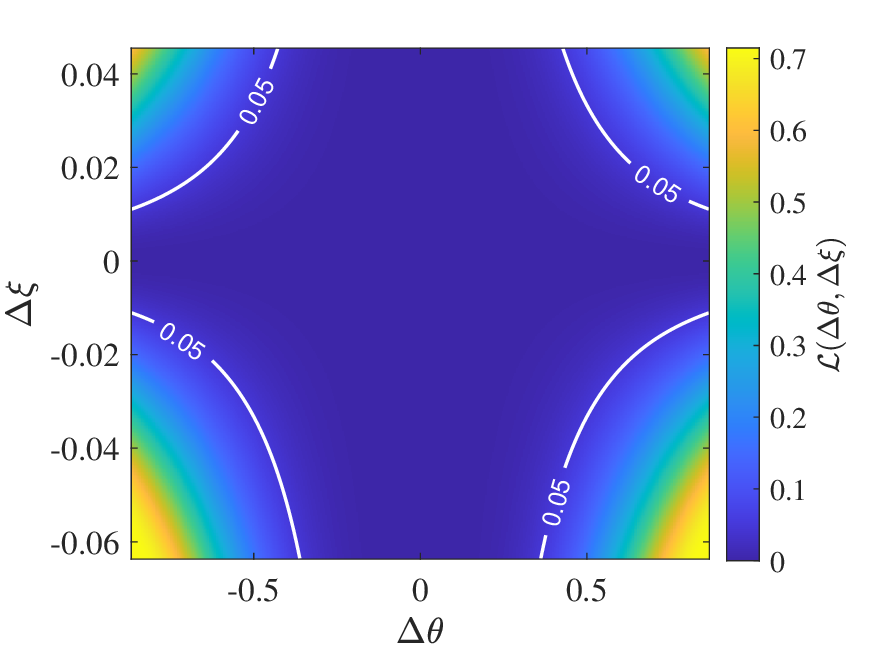}
		\caption{$ \mathcal{L}(\Delta\theta, \Delta\xi) $ versus angle deviation and inverse range deviation when $N=256$, $f=30$ GHz, $\theta_0 = 0$, and $\xi_0 = 1/15$ with $[\theta_{\min}, \theta_{\max}]=[-\frac{\sqrt{3}}{2},\frac{\sqrt{3}}{2}]$ and $[\xi_{\min}, \xi_{\max}] = [ \frac{1}{Z_{\rm Rayl}},  \frac{1}{Z_{\rm Fres}}]$.} \label{Fig:NFTaylor}
		\vspace{-14pt}
	\end{figure}

	By applying the first-order Taylor expansion in {\bf Lemma~\ref{Lem:approx_CSV}}, the phase of near-field CSV is approximated as a linear function w.r.t. $\Delta \theta$ and $\Delta \xi$, which leads to a simplified CSV expression in~\eqref{Exp:approx_CSV} by neglecting the second-order terms.
	To evaluate the approximation accuracy of  $\mathbf{a}_{\rm NF}^{(\rm app)}$ in~\eqref{Exp:approx_CSV}, we first make the following definition.
	\begin{definition}[Beamforming gain loss]\rm Given the exact near-field CSV $\mathbf{a}_{\rm NF}(\theta,\xi)$ and its approximated near-field CSV $\mathbf{a}_{\rm NF}^{(\rm app)}(\theta_0 +\Delta \theta,\xi_0 + \Delta \xi )$ in~\eqref{Exp:approx_CSV}, the beamforming gain loss is defined as 
		\begin{align}
			\mathcal{L}(\Delta\theta, \Delta\xi)  \triangleq 1 - \left| \mathbf{a}_{\rm NF}^H \mathbf{a}_{\rm NF}^{(\rm app)} \right|.
		\end{align}
	\end{definition}
	The defined beamforming gain loss characterizes the discrepancy between the exact and approximated near-field CSVs.  Since the quadratic phase terms are neglected,  such a first-order expansion is accurate only when both $|\Delta \theta|$ and $|\Delta \xi|$ are relatively small. When these deviations become sufficiently large, the second-order terms are no longer negligible, resulting in a noticeable mismatch between the exact and approximated CSVs. 
	To quantify the approximation accuracy, we plot in Fig.~\ref{Fig:NFTaylor} $ \mathcal{L}(\Delta\theta, \Delta\xi) $ versus the deviations in the angle and inverse range.
	It is observed that $\mathcal{L}(\Delta\theta, \Delta\xi)$ increases with the increasing angle and/or inverse-range
	 $\mathcal{L}(\Delta\theta, \Delta\xi)$, indicating that the Taylor approximation becomes less accurate. Notably, when, e.g., $|\Delta \theta |\le \theta_{\rm th}$ (with $\theta_{\rm th}$ serving as a conservative criterion to guarantee negligible beamforming gain loss), $\mathcal{L}(\Delta\theta, \Delta\xi)$ remains relatively small (e.g., $\mathcal{L}(\Delta\theta, \Delta\xi)\le 0.05$).\footnote{This work focuses on scenarios with small angular deviations, i.e., $\theta_{\rm th} =|\Delta \theta|\le 0.2$.  For scenarios with larger angular deviations, the original target region can be partitioned into multiple sub-regions, within which the proposed FT-based beam coverage design can be applied to ensure efficient coverage across the entire target region. This extension will be discussed in Section~\ref{Sec:Lad}.} This indicates that the first-order Taylor approximation provides a sufficiently accurate representation of the near-field CSV within this regime.
	
	{\bf Lemma~\ref{Lem:approx_CSV}} implies that, when the largest angle deviation is small enough (e.g., $|\Delta \theta |\le 0.2$), the phase of near-field CSV can be approximated as a linear function w.r.t. $\Delta \theta$ and $\Delta \xi$, thereby simplifying the expression of the approximated CSV in~\eqref{Exp:approx_CSV}. As such, the beamforming gain at $(\theta,\xi)$ (i.e.,~$| \mathbf{a}_{\rm NF}^H(\theta,\xi) \mathbf{w}_{\rm NF}|$) can be approximated as
	\begin{align}\label{Exp:NF_Fourier}
		\!\big| \mathbf{a}_{\rm NF}^H(\theta,\xi) \mathbf{w}_{\rm NF}\big|	&\approx \Big| \frac{1}{\sqrt{N}}\sum_{n=1}^{N} w_{{\rm NF},n}  e^{-\jmath \frac{2\pi}{\lambda}(\varphi_{n} + \zeta_{n}^{(\theta)}\Delta \theta + \zeta_{n}^{(\xi)}\Delta \xi )} \Big|  \nonumber \\
		&=\Big| \frac{1}{\sqrt{N}} \sum_{n=1}^{N} v_{{\rm NF},n} 
		e^{-\jmath \frac{2\pi}{\lambda}\zeta_{n}^{(\theta)}\Delta \theta} 
		e^{-\jmath \frac{2\pi}{\lambda}\zeta_{n}^{(\xi)}\Delta \xi  }\Big| \nonumber \\
		&\triangleq \big|f_{\rm NF}(\Delta \theta,\Delta \xi;\theta_{0},\xi_{0},N) \big|, 
	\end{align}
	where $v_{{\rm NF},n} = w_{{\rm NF},n}e^{-\jmath \frac{2\pi}{\lambda}\varphi_{n}},\forall n\in\mathcal{N}$. Compared to the far-field beam coverage design, the near-field one can be viewed as a 2D-DFT of the beamforming weight vector $\mathbf{v}_{\rm NF}$, such that the inverse FT can be employed to facilitate near-field beamforming design.

	\subsubsection{FT-based near-field beamforming design}\label{Sec:IV-B}
	Similarly, we select the geometric center of the target region as the reference point $(\theta_{0}, \xi_{0}) $, i.e.,
	\begin{align}
		\theta_{0} = \frac{\theta_{\min}+ \theta_{\max}}{2},
		~\xi_{0} = \frac{\xi_{\min}+ \xi_{\max}}{2}, 
	\end{align} 
	and the corresponding angle and inverse range deviations satisfy
	\begin{align}
			|\Delta \theta|\le \frac{\theta_{\max} - \theta_{\min}}{2} \triangleq \mu,~ 
			|\Delta \xi|\le \frac{\xi_{\max} - \xi_{\min}}{2} \triangleq \nu.
	\end{align}
	To satisfy constraint~\eqref{C:Gain}, we aim to construct a beam pattern with the real-valued beamforming gain for the near-field target region $\mathcal{A}_{\rm NF}$, which is given by
	\begin{equation}\label{Exp:NF_window}
	\!{f}_{\rm NF}(\Delta \theta,\Delta \xi;\theta_{0},\xi_{0},\! N) \!=\! \left\{
		\begin{aligned}
			&\!\gamma_{\rm NF},\;\forall\; |\Delta \theta| \le \mu, |\Delta \xi| \le \nu,\\
			&\! 0,   \quad\;\; \textrm{otherwise}.\\
		\end{aligned}
		\right. 
	\end{equation}

	To achieve the desired beam coverage over the 2D rectangular window in the spatial domain, the beamforming vector $\mathbf{v}_{\rm NF} = [v_{{\rm NF},1}, \ldots, v_{{\rm NF},N}]^{T}$ can be designed following three procedures elaborated below, similar to that adopted in the far-field case.
	Specifically, given the target region $\mathcal{A}_{\rm NF}$ and the reference point $(\theta_{0},\xi_{0})$ with its angle deviation satisfying $|\Delta \theta|\le \theta_{\rm th}$,  the designed transmit beamforming vector for achieving flat beam coverage over $\mathcal{A}_{\rm NF}$ in~\eqref{Exp:NF_window} is given below.
	\begin{mybeamdesign}{\textbf{Proposed near-field beam coverage design}}
		{\bf Step 1} (Add protective zoom): 
		\begin{align}
			\mu^{+} = \mu + {2}/{N}. 
		\end{align}	
		{\bf Step 2} (Weight-shaping design): 
		\begin{align}\label{Exp:NF_weightvec}
			\!\![\tilde{\mathbf{v}}_{\rm NF}]_{n}= \alpha \!\cdot\! 
			\mathrm{sinc}  \Big(\frac{2\mu^{+} \zeta_n^{(\theta)}}{\lambda}\Big)\!\cdot\!
			\mathrm{sinc}\Big(\frac{2\nu\zeta_n^{(\xi)}}{\lambda}\Big),\forall n\in\mathcal{N}.
		\end{align}

	{\bf Step 3} (Beamforming design): 
		\begin{align}\label{Exp:NFBeamvec}
			\tilde{\mathbf{w}}_{\rm NF} = 
			\mathbf{a}_{\rm NF}(\theta_{0},\xi_{0})\odot \tilde{\mathbf{v}}_{\rm NF}.
		\end{align}
	\end{mybeamdesign}\noindent
	Herein, $\tilde{\mathbf{v}}_{\rm NF} \!=\! [\tilde{v}_{{\rm NF},1},\ldots,\tilde{v}_{{\rm NF},N} ]^T$ is the near-field weight-shaping vector, which is given in~\eqref{Exp:NF_weightvec} and proved in Appendix~\ref{App:2},  
	and $\alpha$ is the auxiliary variable introduced to satisfy the transmit power constraint. Similar to the far-field case, a \emph{protective zoom} of width $2/N$ is applied in the angle domain to mitigate the roll-off effect caused by a finite number of antennas.
	In contrast, no protective zoom is imposed in the range domain, which is motivated by the following observation.
	\begin{observation}\rm 
		When the angular coverage is sufficiently large (i.e., $\mu> 2/N$), the resulting beamforming vector will exhibit \emph{steering-beam behavior} along the range domain, which already provides broad range coverage, as characterized in {\bf Proposition~\ref{Pro:defocus}.} 
	\end{observation}
	
	\begin{proposition}[Range defocusing effect]\label{Pro:defocus}\rm
		In the near-field case, consider a beam coverage design that targets only the angle domain (i.e., angle deviation $\mu>0$) with no coverage requirement in the range (inverse-range) domain (i.e., $\nu = 0$). Given a reference point $(\theta_{0},\xi_{0})$, the unconstrained beamforming gain at $(\theta_{0},\xi_{0} + \Delta \xi )$ achieved by the proposed near-field beamforming design can be approximated as
		\begin{align}\label{Exp:NFUAG}
			\big|f_{\rm NF}^{(\rm unc)}(0,\Delta\xi,N)\big| =& \big|\mathbf{a}_{\rm NF}^H(\theta_{0},\xi) \cdot(\mathbf{a}_{\rm NF}(\theta_{0},\xi_{0})\odot \tilde{\mathbf{v}}_{\rm NF})|
			\nonumber \\
			=& \left|\mathbf{a}_{\rm NF}^H(\theta_{0},\xi_{0}+\Delta\xi)\cdot(\mathbf{a}_{\rm NF}(\theta_{0},\xi_{0})\odot \tilde{\mathbf{v}}_{\rm NF} )\right| \nonumber \\ 
			\approx&  \frac{1}{2D} \int_{0}^{1} \Big[ \mathcal{I}(t) + \mathcal{I}(-t) \Big] \; dt,
		\end{align}
		where \begin{align}
			\mathcal{I}(t) & = \sqrt{\frac{\pi}{2|\mathcal{Q}(t)|}} e^{-\jmath \frac{\mathcal{J}^2(t)}{4\mathcal{Q}(t)}} \times \nonumber \\
			&\quad\;\; \begin{cases}
				\big[ \mathcal{G}(z_2(t)) - \mathcal{G}(z_1(t)) \big], & \mathcal{Q}(t) > 0 \\
				\big[ \mathcal{G}^*(z_2(t)) - \mathcal{G}^*(z_1(t)) \big], & \mathcal{Q}(t) < 0
			\end{cases}.
		\end{align}
		Herein, $\mathcal{Q}(t) = \psi + \pi \kappa \varkappa t$ and $\mathcal{J}(t) = \pi \kappa t$,
		where $\kappa = \frac{2\mu}{\lambda}$, $\varkappa=\theta_{0}\xi_{0}$, and
		$\psi = \frac{\pi (1-\theta_{0}^{2})\Delta\xi}{\lambda}$.
		Additionally, 
		$z_1(t) = \sqrt{\frac{2|\mathcal{Q}(t)|}{\pi}} \left( - \frac{D}{2} + \frac{\mathcal{J}(t)}{2\mathcal{Q}(t)} \right)$,
		$z_2(t) = \sqrt{\frac{2|\mathcal{Q}(t)|}{\pi}} \left( \frac{D}{2} + \frac{\mathcal{J}(t)}{2\mathcal{Q}(t)} \right)$, $\mathcal{G}(z) = \mathcal{C}(z) + \jmath\mathcal{S}(z)$, where $\mathcal{C}(z) = \int_{0}^{z} \cos\left(\frac{\pi}{2}t^2\right) dt$ and $\mathcal{S}(z) = \int_{0}^{z} \sin\left(\frac{\pi}{2}t^2\right) dt$ are the Fresnel functions.
	\end{proposition}
	\begin{proof}
		Please refer to Appendix~\ref{App:beamgain}.
	\end{proof}
 
	{\bf Proposition~\ref{Pro:defocus}} characterizes the unconstrained beamforming gain along the range domain under the angle-only beam coverage. While \eqref{Exp:NFUAG} provides a closed-form expression for the unconstrained beamforming gain in the range domain, the relationship between the angular beam-width and the range-domain behavior involves complex Fresnel integrals, making it difficult to obtain useful insights. Therefore, a numerical example is provided to facilitate intuitive understanding.

	\begin{example}\label{Exa:3}\rm 
		\begin{figure}[t]
			\centering
			\includegraphics[width=0.33\textwidth]{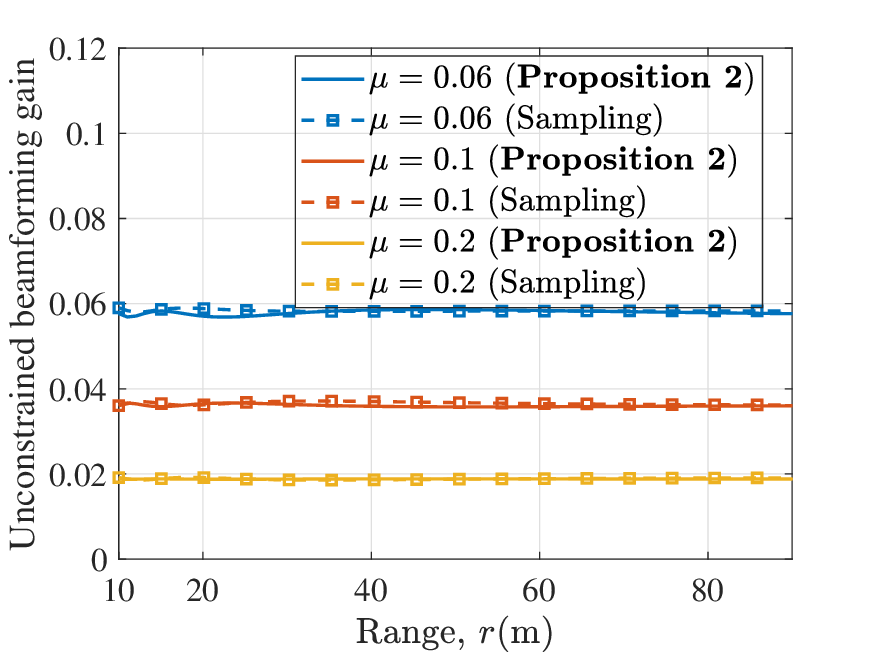}
			\caption{Unconstrained beamforming gain in the range domain  for the near-field case given inverse range deviation $\Delta \xi = 0$, reference point $(0,1/15)$, $N=256$, and $f=30$ GHz under different angle deviations~$\mu$.} \label{Fig:NFdefocusing}
			\vspace{-14pt}
		\end{figure}
		To visualize the range defocusing behavior of {\bf Proposition~\ref{Pro:defocus}}, we present in Fig.~\ref{Fig:NFdefocusing} the unconstrained beamforming gain in the range domain for different angle deviations $\mu$, with a reference point $(0,1/15)$,  $N=256$, and $f=30$ GHz. Several key observations can be drawn.
		\begin{itemize}
			\item First, as the angular deviation $\mu$ increases, a noticeable reduction in the unconstrained beamforming gain is observed. This phenomenon reveals an inherent trade-off that enlarging the angular coverage inevitably comes at the expense of achievable beamforming gain.
			
			\item Second, for different values of the angular deviation $\mu$,  the proposed FT-based beam coverage method and the sampling-based method exhibit comparable beamforming gain along the range domain.
			
			\item Third, the beamforming gain remains nearly flat along the range domain for all considered angular deviations. This observation indicates that, under wide angular coverage, near-field beam-focusing reduces to a range-insensitive beam steering, which is referred to as the near-field \textit{range defocusing} effect.\footnote{To maximize the worst-case beam coverage performance in the near-field case under wide angular coverage, both the proposed and sampling-based methods exhibit the range defocusing effect (see Fig.~\ref{Fig:NFdefocusing}). While for other objectives aiming to achieve an angle-and-range focused beam pattern over a specified region, existing methods, such as the Gerchberg–Saxton (GS)-based algorithm~\cite{LuHBT} and optimization-based methods~\cite{Wang2025,Weng26}, can be employed to design the target beam pattern in the angle and range domains. However, low-complexity design for such scenarios is still challenging and is left for our future work.}
		\end{itemize}
	\end{example}
	
	\begin{remark}[Why defocus in the range domain~?] \rm 
		{\bf Proposition~\ref{Pro:defocus}} and {\bf Example~\ref{Exa:3}} reveal that extending the angular coverage comes at the cost of degraded near-field focusing capability. Specifically, when a flat angular coverage is enforced, the resulting power allocation across the antenna aperture approaches a $\mathrm{sinc}$-shaped profile, with the transmit energy predominantly concentrated around the aperture center. Under this condition, only a subset of the central antenna elements effectively contribute to beam coverage, while the remaining antennas are actually inactive in the beam shaping. As a result, the effective aperture size is smaller than the physical aperture, leading to the observed beam defocusing in the range domain.
	\end{remark}

	\begin{remark}[Advantages of proposed method]\rm
		In contrast to existing beam coverage designs (e.g., sampling-based methods) that rely on dense spatial sampling and iterative optimization over a large set of angular or range grid points, the proposed approach admits a closed-form beamforming solution that is directly parameterized by the target region. Specifically, for the sampling-based scheme with a sampling number of $S$, its computational complexity is in the order of $\mathcal{O}\left(I \sqrt{2(S+1)}\cdot \log(1/\epsilon)\left( \left( S+2\right) N^3\right) \right)$, where $I$ is the number of SCA iterations and $\epsilon$ is a predefined accuracy~\cite{MyTPM}. 
		By avoiding iterative procedures and exhaustive grid searches, the resulting computational complexity is significantly reduced, while the achieved worst-case beamforming gain is comparable to that of sampling-based benchmarks (see numerical results in Section~\ref{Sec:NR}). 
	\end{remark}

	\begin{remark}[Far-field versus near-field cases]\rm
		The proposed beam coverage design method for both far-field and near-field scenarios shares a similar framework based on FT, while the main differences arise from the dimensionality of the target region and the underlying spatial-frequency mapping.
		
		\begin{itemize}
			\item \textbf{Far-field case:}  
			The beam coverage design reduces to a one-dimensional (1D) problem in the angular domain, characterized solely by the deviation $\Delta\theta$. 
			In this case, the phase of CSV is inherently linear w.r.t. the spatial-frequency variable $u_n \Delta\theta$, which directly establishes the 1D FT relationship in \eqref{Exp:FF_Fourier}. 
			As a result, the corresponding beamforming weights admit a closed-form $\mathrm{sinc}(\cdot)$ structure, as given in \eqref{Exp:FF_weightvec}.
			
			\item \textbf{Near-field case:}  
			The	beam coverage design becomes a 2D problem over the joint angle and range domains, parameterized by $(\Delta\theta,\Delta\xi)$. 
			Due to spherical-wave propagation, the phase of near-field CSV in \eqref{Exp:NFCSV} is inherently nonlinear.
			To make the beam coverage design problem more tractable, a first-order Taylor approximation around the reference point is applied, as presented in {\bf Lemma~\ref{Lem:approx_CSV}}, which linearizes the phase w.r.t. the deviations $(\Delta\theta,\Delta\xi)$. 
			This approximation yields an approximate 2D FT relationship, as shown in~\eqref{Exp:NF_Fourier}. As such, it leads to a separable beamforming structure composed of the product of two $\mathrm{sinc}(\cdot)$ functions in \eqref{Exp:NF_weightvec}, corresponding to a rectangular target region in the angle and range domains.
		\end{itemize}
	
		Notably, the far-field design can also be viewed as a special case of the near-field case in~\eqref{Exp:NFBeamvec} by 
		setting $\xi_{0}=0$ and $\Delta \xi \!=\!0$.
	\end{remark}
	\subsection{Discussions and Extensions}\label{Sec:dis}
	
	\subsubsection{\underline{\bf{\emph{Multi-region beam coverage}}}}
	The proposed low-complexity FT-based beamforming framework can be easily extended to the multi-region beam coverage scenario, where simultaneous beam coverage over multiple spatially separated regions is required.
	
	Specifically, for the far-field case, consider a set of $K$ non-overlapping target regions $\{\mathcal{A}_{{\rm FF},k}\}_{k=1}^{K}$ (denoted by $\mathcal{K}\triangleq\{1,2,\ldots,K\}$), each characterized by a reference angle $\theta_{0,k}$ and an angle deviation $\mu_k$. To ensure negligible inter-beam interference, these regions are assumed to be sufficiently separated in the angle domain, satisfying
	\begin{align}
		| \theta_{0,i} - \theta_{0,j} | \ge \mu_i + \mu_j + {8}/{N},  \;\forall i \neq j \in \mathcal{K},
	\end{align}
	where the margin $\frac{8}{N}$ guarantees that the roll-off regions of adjacent beams do not overlap, thereby preserving approximate beam orthogonality.

	For each region $\mathcal{A}_{{\rm FF},k},k\in\mathcal{K}$, the associated beamforming vector $\tilde{\mathbf{w}}_{{\rm FF},k}$ can be obtained by using the proposed beam coverage design in Section~\ref{Sec:FFcase}. As such, the transmit vector is constructed as a weighted superposition of $\{\tilde{\mathbf{w}}_{{\rm FF},k}\}$, i.e.,
	\begin{align}
		\mathbf{w}_{\rm FF}^{(\rm MB)} = \sum_{k\in\mathcal{K}} \beta_k \tilde{\mathbf{w}}_{{\rm FF},k},
	\end{align}
	where $\{\beta_k\}_{k=1}^{K}$ are scaling coefficients for power normalization and flexible gain allocation across different beams.
	
	The proposed multi-region beam coverage design can be readily extended to the near-field case to reduce the associated beam training overhead, with the details omitted for brevity.

	\subsubsection{\underline{\bf{\emph{Large angle deviation}}}}\label{Sec:Lad}	
	In near-field scenarios with large angular deviations, directly applying the approximated near-field CSV over the entire target region may incur non-negligible approximation errors. To address this issue, the target region can be partitioned into multiple sub-regions, within which the CSV approximation remains locally accurate and the proposed beam coverage design can be applied.
		
	Specifically, consider a near-field target region $\mathcal{A}_{\rm NF}$ with $|\Delta \theta| > \theta_{\rm th}$.  We first partition the original region $\mathcal{A}_{\rm NF}$ into $M$ sub-regions $\{	\mathcal{B}_{{\rm NF},m} \triangleq \{(\theta,\xi) \mid \theta \in \Theta_m, \xi \in \Xi \}\}_{m=1}^{M}$(indexed by $\mathcal{M} \triangleq \{1,2,\ldots,M\}$),  with $\Theta_m \triangleq [\theta_{m,\min}, \theta_{m,\max}]$, such that the maximum angle deviation within each sub-region is constrained by $|\Delta \theta| \le \theta_{\rm th} $. To ensure flat and continuous beam coverage over the near-field target region $\mathcal{A}_{\rm NF}$, the sub-regions are designed to satisfy the following conditions:
	(i) the union of all sub-regions fully covers the original target region,
	(ii) the near-field CSV approximation within each sub-region is accurate enough, and
	(iii) a fixed angular gap of $4/N$ is maintained between adjacent sub-regions. Based on the above, the resultant beamforming vector is constructed as
	\begin{align}
		\mathbf{w}_{\rm NF}^{(\rm LAD)} = \sum_{m\in \mathcal{M}} \beta_m \tilde{\mathbf{w}}_{{\rm NF,m}},
	\end{align}
	where $ \tilde{\mathbf{w}}_{{\rm NF,m}} $ denotes the beamforming vector designed for the $m$-th sub-region and $\{\beta_m\}_{m=1}^{M}$ are scaling coefficients introduced to satisfy the transmit power constraint.
	 
	\subsubsection{\underline{\bf{\emph{Analog beamforming}}}}
	For analog beamforming architectures with constant-modulus constraints, i.e., $|[\mathbf{w}]_n| = 1/\sqrt{N},~\forall n$, phase-only beam broadening can be achieved by exploiting the FT property that a linear frequency modulation (LFM) signal in the time domain yields a rectangular magnitude spectrum in the frequency domain.
	
	Specifically, the phase profile across the array aperture can be designed as follows~\cite{LFM_broad}
	\begin{align}
		\Phi(u_{n}) = \frac{2\pi}{\lambda}
		\big(\underbrace{\theta_0 u_{n}}_{\text{linear phase}}
		+ \underbrace{\eta u_{n}^2}_{\text{quadratic phase}}\big),
	\end{align}
	where $\eta$ is the quadratic phase coefficient introduced for 
	controlling the beamwidth. Based on the above, the corresponding beamforming vector is given by
	\begin{align}
		[\mathbf{w}_{\rm NF}^{(\rm PA)}]_n = \frac{1}{\sqrt{N}} e^{\jmath \frac{2\pi}{\lambda} (  \theta_0 u_n + \eta u_n^2 )},\forall n\in\mathcal{N},
	\end{align}
	where $\eta = \frac{1}{2 D^2} ( 2D \varpi + \lambda + \sqrt{\lambda(4D \varpi + \lambda)} ) $ with $\varpi = \mu \sqrt{1 - \frac{\theta_0^2}{1-\mu^2}}$.
	The above design follows the LFM-inspired quadratic phase modulation method in~\cite{LFM_broad}, with detailed derivation omitted for brevity.

	For the near-field case, quadratic phase modulation can also be exploited to broaden angular coverage while satisfying the constant-modulus constraint. However, unlike the far-field case, near-field beamforming requires joint coverage in both the angle and range domains. 
	As a result, although phase-only modulation enables effective angular broadening, extending the beam coverage in the range domain by using the LFM-inspired phase-only design remains nontrivial, which is left for our future work.

	\section{Numerical Results}\label{Sec:NR}
		\begin{figure*}[t]
		\centering
		\begin{subfigure}[b]{0.32\linewidth}
			\includegraphics[width=1\linewidth]{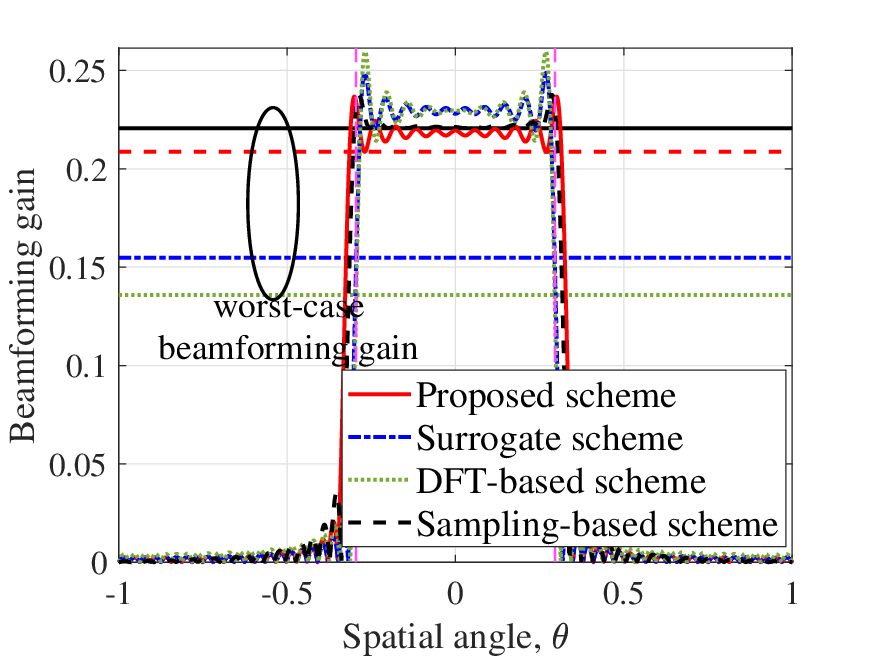}
			\caption{Beam pattern.}
			\label{Fig:FFpattern}
		\end{subfigure}
		\begin{subfigure}[b]{0.32\linewidth}
			\includegraphics[width=1\linewidth]{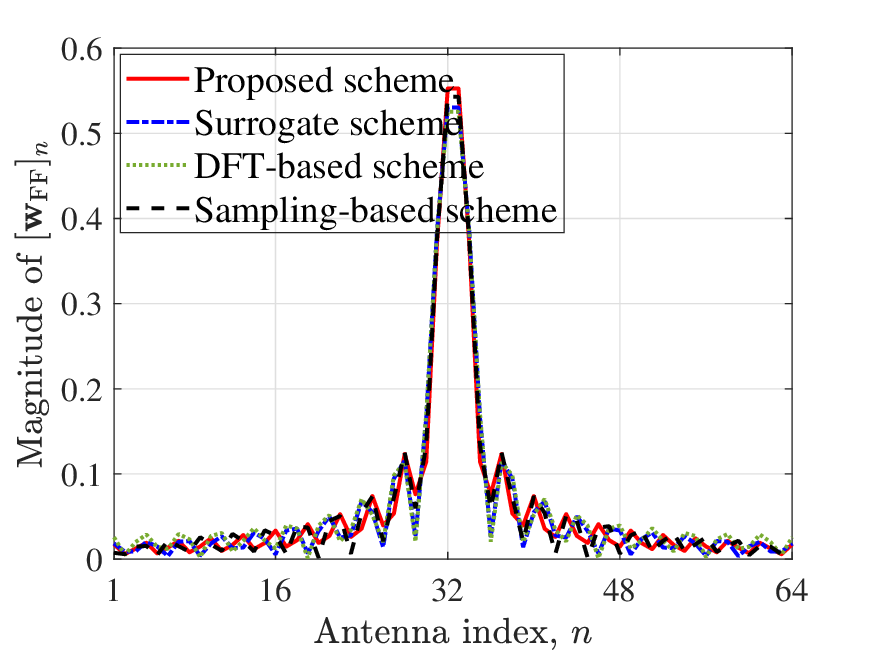}
			\caption{Magnitude of $[\mathbf{w}_{\rm FF}]_{n}$.}
			\label{Fig:FFweightshaping}
		\end{subfigure}
		\begin{subfigure}[b]{0.32\linewidth}
			\includegraphics[width=1\linewidth]{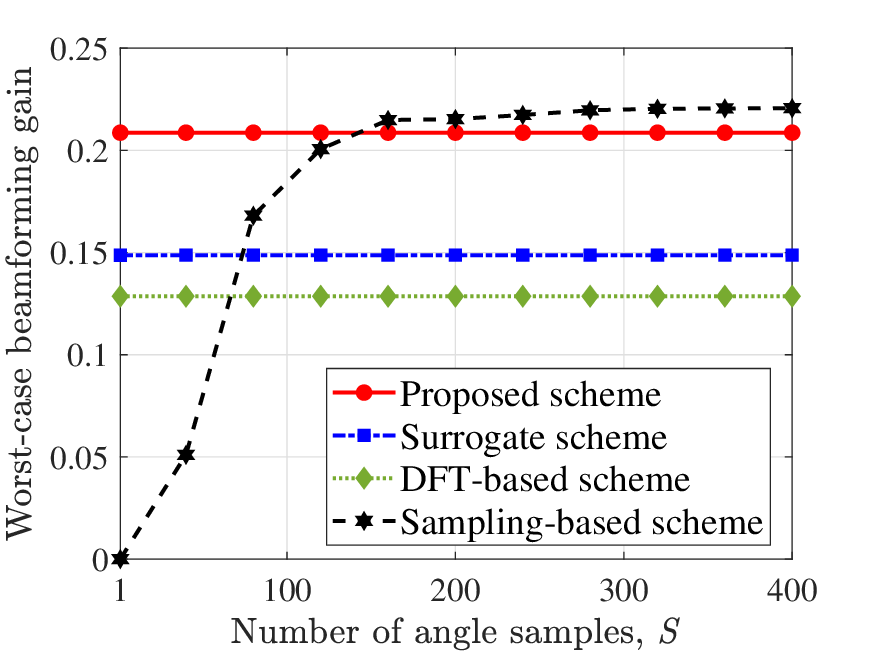}
			\caption{Worst-case beamforming gain.}
			\label{Fig:FFgain}
		\end{subfigure}
		\caption{Beam coverage design for the far-field case.}
		\label{Fig:FFBCD}
		\vspace{-10pt} 
	\end{figure*}
	\begin{figure*}[t]
		\centering
		\begin{subfigure}[b]{0.32\linewidth}
			\includegraphics[width=1\linewidth]{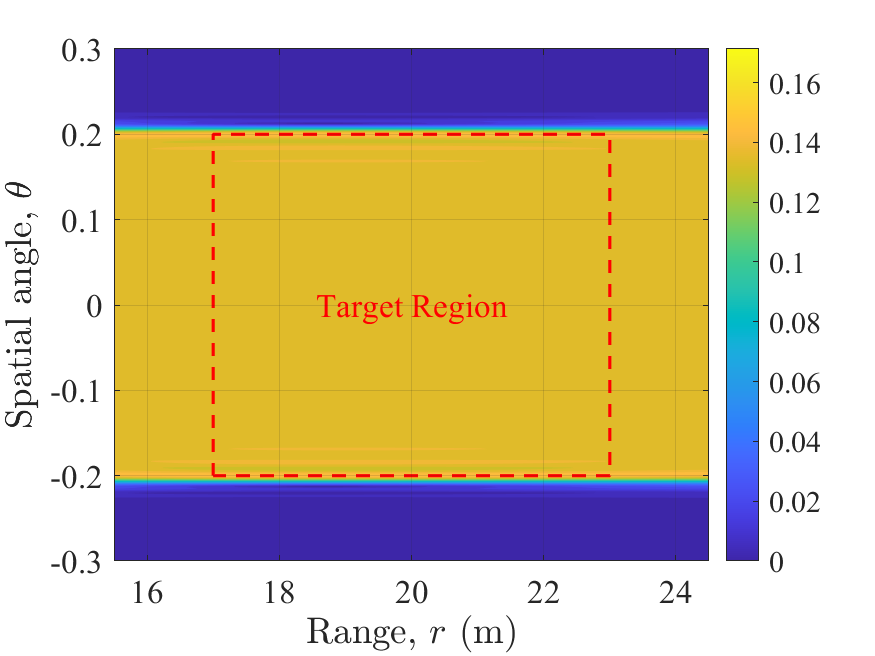}
			\caption{Beam pattern.}
			\label{Fig:NFpattern}
		\end{subfigure}
		\begin{subfigure}[b]{0.32\linewidth}
			\includegraphics[width=1\linewidth]{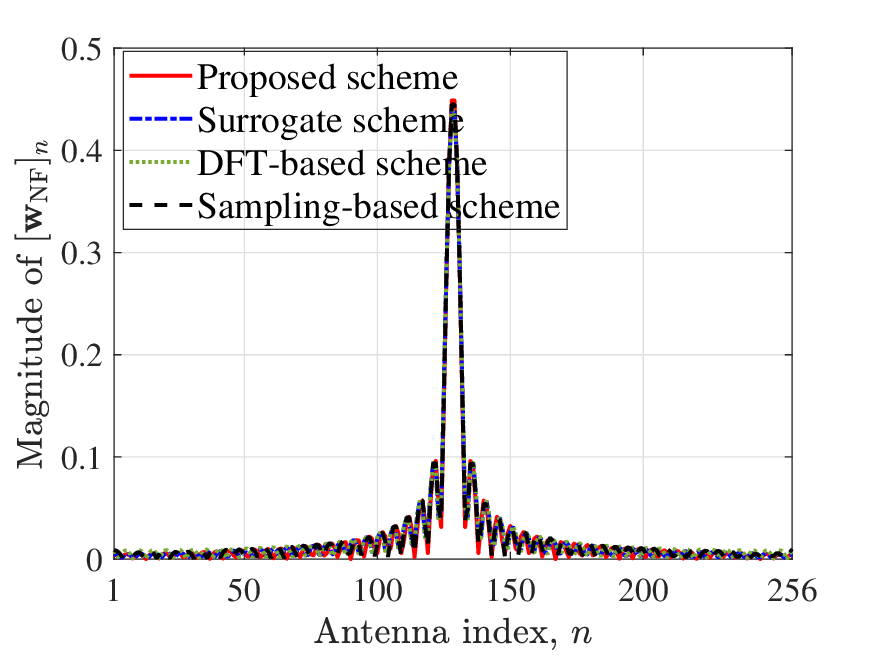}
			\caption{Magnitude of $[\mathbf{w}_{\rm NF}]_{n}$.}
			\label{Fig:NFweightshaping}
		\end{subfigure}
		\begin{subfigure}[b]{0.32\linewidth}
			\includegraphics[width=1\linewidth]{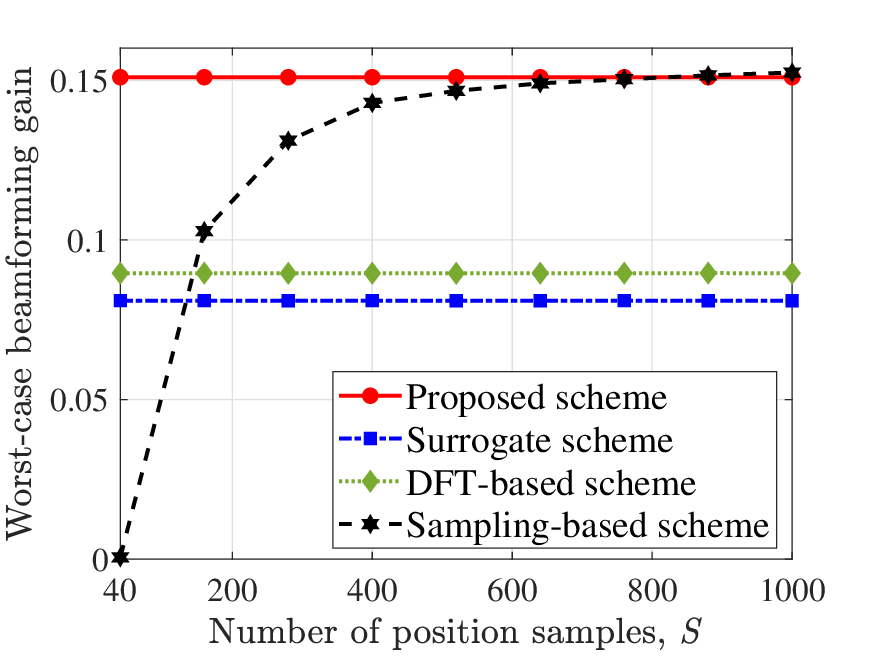}
			\caption{Worst-case beamforming gain.}
			\label{Fig:NFgain}
		\end{subfigure}
		\caption{Beam coverage design for the near-field case.}
		\label{Fig:NFBCD}
		\vspace{-10pt} 
	\end{figure*}
	Numerical results are presented in this section to demonstrate the coverage performance of proposed low-complexity beamforming designs. For the far-field case, the BS is equipped with $64$ antennas operating at $f = 30$ GHz frequency band, targeting the beam coverage within a spatial angle interval $\Theta = [-0.3, 0.3]$, which corresponds to a physical angle interval $[-17.46^\circ,17.46^\circ]$.
	For the near-field case, an XL-array comprising $N = 256$ antennas at $f = 30$ GHz is considered, which aims to achieve 2D beam coverage over $\Theta = [-0.15, 0.15]$ and $\Xi = [1/23, 1/17]$ (equivalently a physical angle interval $[-8.63^\circ,8.63^\circ]$ and range interval $[17, 23]$ meter~(m)).
	The following benchmark schemes are considered for performance comparison. 
	\begin{itemize}
		\item \emph{Surrogate scheme}:
		For this scheme, the beamforming vector is designed by using the inverse FT without adding the protective zoom.
		
		\item \emph{Sampling-based scheme}:
		For this scheme, the target region $\mathcal{A}_{\chi},\chi\in\{\rm FF,\rm NF\}$ is uniformly discretized with a sampling number of $S$. As such, the original problem is reformulated and solved by addressing $S$ second-order cone (SOC) constraints~\cite{LiuSOC}.
		
		\item \emph{DFT-based scheme}: 
		For this scheme, the beamforming vector is obtained by summing all DFT codewords within the target angular interval, followed by power normalization to satisfy the transmit power constraint~\cite{Xu}.
	\end{itemize}

	\subsection{Computational Complexity}
	First, in Table~\ref{tab:CC}, we compare the average runtime of different schemes. As expected, the sampling-based scheme takes the longest runtime, which increases substantially in the near-field case due to the larger number of antennas and the resulting high-dimensional signal processing. In contrast, the proposed scheme features a much lower computational complexity, leading to an approximately five orders-of-magnitude reduction in runtime compared with the sampling-based scheme (with $S=20$) in both the far-field and near-field cases. Moreover, although the DFT-based scheme avoids iterative optimization, it still incurs a higher runtime than the proposed scheme, due to the computational overhead associated with large-scale matrix operations.
	\begin{table}[!t]
	\centering
	\caption{Runtime comparison among different schemes.}
	\label{tab:CC}
	\renewcommand{\arraystretch}{1.2}
	\begin{tabular}{|l|c|c|}
		\hline
		\multirow{2}{*}{\textbf{Scheme}} & \multicolumn{2}{c|}{\textbf{Runtime (ms)}} \\ 
		\cline{2-3} 
		& \textbf{Far-field case} & \textbf{Near-field case} \\
		\hline
		Proposed scheme       & 0.0074  & 0.0292 \\
		\hline
		Surrogate scheme      & 0.0073  & 0.0291 \\
		\hline
		Sampling-based scheme $(S=20)$ & 884.5942 & 1275.066 \\
		\hline
		DFT-based scheme      & 0.0537 & 0.3746 \\
		\hline
	\end{tabular}
	\vspace{-14pt}
	\end{table}
	
	\subsection{Beam Coverage Design for Far-field Case}

	To evaluate the performance of proposed beam coverage design in the far-field case, we present in Figs.~\ref{Fig:FFBCD}(a)--\ref{Fig:FFBCD}(c) the beam pattern, the magnitude of $[\mathbf{w}_{\mathrm{FF}}]_n$, and the worst-case beamforming gain for different schemes, respectively. As shown in Fig.~\ref{Fig:FFBCD}(a), the proposed scheme achieves a beam pattern that is comparable to the sampling-based scheme within the desired target region, with the roll-off effect occurring outside the target angular region, thereby achieving a satisfactory worst-case beamforming gain. In contrast, both the surrogate and DFT-based schemes exhibit significant roll-off at the boundaries of the target region, resulting in degraded beamforming gain. Additionally, Fig.~\ref{Fig:FFBCD}(b) shows the magnitude of $[\mathbf{w}_{\mathrm{FF}}]_n$ for different schemes. The energy distribution across the antenna domain for all schemes exhibits similar $\sinc(\cdot)$-like shapes, as given in~\eqref{Exp:FF_weightvec}. This indicates that, in the fully digital beamforming architecture, the beam coverage design is primarily determined by the power control across the antenna domain. Consequently, the proposed scheme, which considers only weight shaping, achieves a worst-case beamforming gain close to that of the sampling-based scheme. Furthermore, we plot in Fig.~\ref{Fig:FFBCD}(c) the worst-case beamforming gain versus the number of angular samples. It is observed that the proposed scheme, which does not rely on dense sampling, achieves a worst-case beamforming gain comparable to that of the sampling-based scheme; while the latter one requires a significantly large number of samples (e.g., $S > 100$) and hence demanding computational complexity.

	\subsection{Beam Coverage Design for Near-field Case}
	
	Next, we investigate the beam coverage performance of proposed scheme in the near-field case. It is observed that by employing the proposed beam coverage design, efficient beam coverage is achieved in both the angle and range domains, as shown in Fig.~\ref{Fig:NFBCD}(a). Moreover, unlike conventional near-field beam focusing, the angular coverage in the near-field case induces beam defocusing in the range domain, leading to a flat steering beam along the range domain, as described in \textbf{Proposition~\ref{Pro:defocus}}. Additionally, a similar $\sinc(\cdot)$-like power distribution of $|[\mathbf{w}_{\rm NF}]_{n}|$ across the antenna domain can be observed in Fig.~\ref{Fig:NFBCD}(b), which demonstrates that the proposed scheme is capable of flexible beam coverage design without leveraging the phase DoFs. In Fig.~\ref{Fig:NFBCD}(c), we compare the worst-case beamforming gain for different schemes. The sampling-based method improves as the number of samples increases, owing to the finer traversal of the target region, which leads to higher beamforming gain but comes at the cost of significantly increased computational complexity. Furthermore, compared to the far-field case, more sampling points are required in the near-field case to achieve a worst-case beamforming gain similar to that of the proposed scheme.

	\subsection{Large Angle Deviation Scenario}
\begin{figure}[t]
		\centering
		\begin{subfigure}[b]{0.49\linewidth}
			\includegraphics[width=1\linewidth]{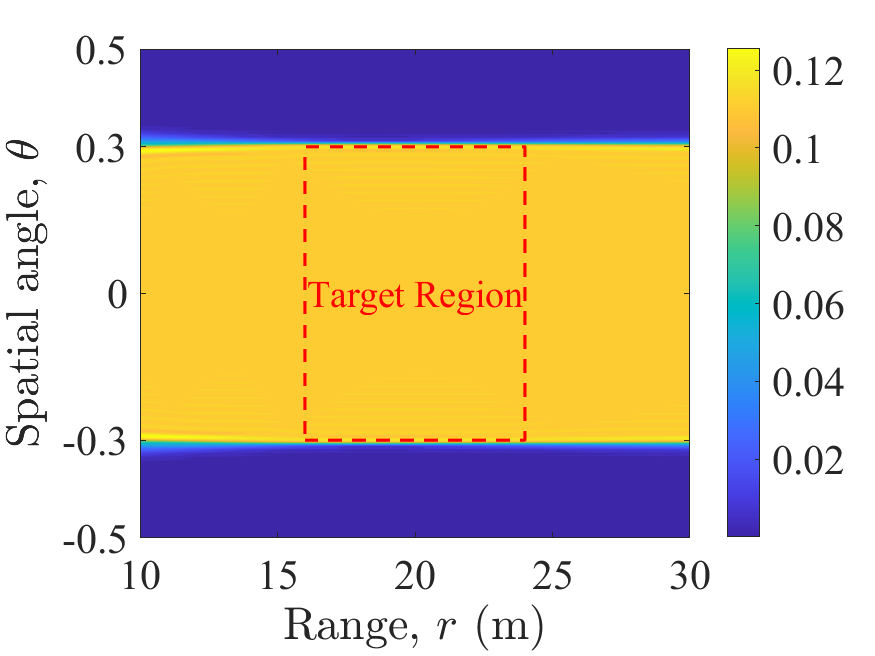}
			\caption{Beam pattern in the angle-range domain.}
			\label{Fig:NFLADpattern1}
		\end{subfigure}
		\hspace{-2pt}
		\begin{subfigure}[b]{0.49\linewidth}
			\includegraphics[width=1\linewidth]{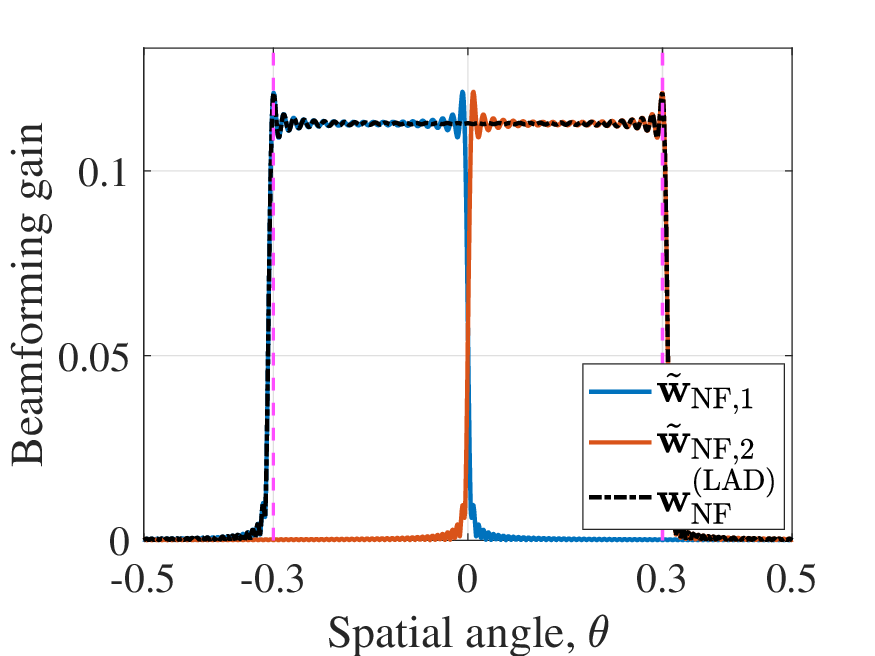}
			\caption{Beam pattern in the angle domain.}
			\label{Fig:NFLADpattern2}
		\end{subfigure}
		\caption{Beam patterns of proposed scheme for large angle deviation scenario.}
		\label{Fig:LAD}
		\vspace{-10pt} 
	\end{figure}

	For near-field cases with large angle deviations (i.e., $\mathcal{A}_{{\rm NF}} = \{(\theta,\xi ) \mid \theta \in [-0.3, 0.3], \xi \in [1/24,1/16] \}$), we present in Fig.~\ref{Fig:LAD} the beam patterns of proposed scheme for large angle deviation scenario. It is observed in Fig.~\ref{Fig:LAD}(a) that the proposed scheme maintains effective beam coverage even when the maximum angular deviation $\mu$ exceeds the predefined threshold (e.g., $\theta_{\rm th} = 0.2$).
	In such scenarios, the global accuracy of the approximated near-field CSV, $\mathbf{a}_{\rm NF}^{(\rm app)}$, typically deteriorates. However, the proposed scheme effectively addresses this issue by partitioning the original target region into multiple sub-regions, within which $\mathbf{a}_{\rm NF}^{(\rm app)}$ remains locally accurate.
	As a result, by judiciously selecting the reference points and designing the angular spacing between adjacent sub-regions, the coherent superposition of these beams designed for each sub-region yields a uniform and flat-top beamforming gain across the entire target region, as shown in Fig.~\ref{Fig:LAD}(b).
	
	\vspace{-6pt}
	\section{Conclusions}\label{Sec:Con}
	
	In this paper, we proposed a low-complexity beam coverage design for multi-antenna systems in both far-field and near-field cases. By exploiting the FT relationship between the antenna and spatial domains, we showed that far-field angular coverage could be efficiently realized via an inverse FT. To mitigate the beamforming loss caused by the roll-off effect of the surrogate scheme, which directly truncates the infinite-length sequence, a roll-off-aware beam coverage design was developed by judiciously introducing a protective zoom, ensuring a flat beamforming-gain profile within the target angular region.
	We then extended this FT-based beam coverage design to the near-field case by applying a first-order Taylor approximation to the phase of near-field CSV, enabling a 2D inverse FT for joint angle and range beam coverage. Moreover, an inherent range defocusing effect in the near-field beam coverage was observed when wide angular coverage was enforced. Numerical results demonstrated that the proposed FT-based designs achieve comparable beam coverage performance with those of conventional sampling-based optimization methods, while significantly reducing the computational complexity.

	\begin{appendices}
		\section{}\label{App:IdealWS}	
			The vector $\{\bar{s}_{{\rm FF},n}\}$ can be obtained by using the inverse FT as follows
			\begin{align}
				&\bar{s}_{{\rm FF},n} = 
				\int_{\mathbb{R}} {f}_{\rm FF}(\Delta \theta;\theta_{0},N)\,
				e^{\jmath\frac{2\pi}{\lambda}(u_{n}\Delta \theta )}
				\,d(\Delta\theta) \nonumber \\
				&\quad \; \quad = \int_{-\mu\le \Delta\theta\le \mu}
				{\gamma_{\rm FF}}\,
				e^{\jmath\frac{2\pi}{\lambda}(u_{n}\Delta \theta )}
				\,d(\Delta\theta) \nonumber \\
				&=  {\gamma_{\rm FF}}
				\int_{-\mu}^{\mu}  e^{\jmath \frac{2\pi}{\lambda} u_{n} x } dx  =  2 \mu {\gamma_{\rm FF}} \sinc\big(\frac{2 u_{n}\mu}{\lambda}\big), n\in \mathcal{Z},
			\end{align}
			where $\sinc(a)$ is defined below~\eqref{Exp:FF_weightvec}, thus completing the proof.

		\section{}\label{App:UAG}	
		The unconstrained beamforming gain can be rewritten as 
		\begin{align}\label{Exp:fin_UAG}
			&\left| {f}_{\rm FF}^{(\rm unc)}(\Delta \theta,N)\right|  = \left| \mathbf{a}_{\rm FF}^{H}(\theta_0 + \Delta \theta)\cdot \big(\mathbf{a}_{\rm FF}(\theta_{0}) \odot \mathbf{v}_{\rm FF}\big)\right| \nonumber \\
			=&	\left|  \sum_{n=1}^{N} \left(\frac{1}{\sqrt{N}} e^{-\jmath\frac{2\pi}{\lambda}u_{n}(\theta_{0}+\Delta \theta)} \right)  \left(\frac{1}{\sqrt{N}} e^{\jmath\frac{2\pi}{\lambda}u_{n}\theta_{0}} \right) [\mathbf{v}_{\rm FF}]_{n} \right| \nonumber \\  
			=&\frac{1}{N} \left| \sum_{n=1}^{N} [\mathbf{v}_{\rm FF}]_{n}  e^{-\jmath\frac{2\pi}{\lambda}u_{n}\Delta \theta}   \right|. 
		\end{align}
		Based on $	{v}_{{\rm FF},n}^{} = \bar{v}_{{\rm FF},n}^{} \cdot m_{n}$, the unconstrained beamforming gain in~\eqref{Exp:fin_UAG} can be re-expressed as follows by extending the summation domain from the finite set $\mathcal{N}$ to the infinite integer set $\mathcal{Z}$
		\begin{align}\label{Exp:infin_UAG}
			\left| {f}_{\rm FF}^{(\rm unc)}(\Delta \theta,N)\right|\!= \! \frac{1}{N} \left| \sum_{n=-\infty}^{+\infty} \!\left( \bar{v}_{{\rm FF}, n} \cdot m_n \right) e^{-\jmath \frac{2\pi}{\lambda} u_n \Delta \theta} \right|, 
		\end{align}
		which represents the DFT of element-wise product of two sequence $\{\bar{\mathbf{v}}_{{\rm FF},n}\}$ and $\{m_{n}\}$. According to the convolution theorem, the multiplication in the spatial domain corresponds to the convolution in the spatial frequency. Specifically, the FT of ideal weight-shaping vector $\bar{\mathbf{v}}_{\rm FF}$ and rectangular mask $\mathbf{m}$ are given by
		\begin{subequations}
			\begin{align}
				& \mathcal{F} \big\{ \bar{\mathbf{v}}_{\rm FF} \big\} (\Delta \theta) \triangleq f_{\rm \bar{v}}(\Delta \theta) = 
				\begin{cases} 
					\frac{1}{2\mu}, & \text{if } |\Delta \theta| \le \mu, \\
					0, & \text{otherwise}.
				\end{cases} \\
				& \mathcal{F} \big\{ \mathbf{m} \big\} (\Delta \theta) \triangleq \Omega(\Delta \theta,N)  = e^{-\jmath\frac{\pi\Delta \theta(N+1)}{2}}\frac{\sin(\frac{N \pi \Delta \theta}{2})}{\sin(\frac{\pi \Delta \theta}{2})}.
			\end{align}
		\end{subequations}
		As such, the unconstrained beamforming gain in~\eqref{Exp:infin_UAG} can be written as follows in a convolution form
		\begin{align}
			\left| {f}_{\rm FF}^{(\rm unc)}(\Delta \theta,N)\right|= & \frac{1}{N} \left|   f_{\rm \bar{v}}(\Delta \theta)  \circledast   \Omega(\Delta \theta,N) \right|  \nonumber \\
			=& \frac{1}{N} \left| \int_{-\infty}^{+\infty} f_{\rm \bar{v}}(x) \Omega(\Delta \theta - x,N) \, dx \right| \nonumber \\
			=& \frac{1}{N} \left| \int_{-\mu}^{\mu} \frac{1}{2\mu} \Omega(\Delta \theta - x,N) \, dx \right| \nonumber \\
			=& \frac{1}{2\mu N} \left| \int_{-\mu}^{\mu} N \frac{\sin\big(\frac{N \pi (\Delta\theta - x)}{2}\big)}{\frac{N \pi (\Delta\theta - x)}{2}} \, dx \right|,
		\end{align}
		which completes the proof.

		\section{}\label{App:Sifun}	
		Let $t = \frac{N\pi}{2}(\Delta \theta-x)$. The unconstrained beamforming gain $|{f}_{\rm FF}^{(\rm unc)}(\Delta \theta,N)|$ in~\eqref{Exp:Conv} can be rewritten as 
		\begin{align}\label{Exp:Conv_expansion} 
			&\l|f_{\rm FF}^{(\rm unc)}\r| = \int_{t_{1}}^{t_{2}} N \frac{\sin(t)}{t} \left(-\frac{2}{N \pi}\right) dt 
			\; = \frac{2}{\pi} \int_{t_{2}}^{t_{1}} \frac{\sin(t)}{t} dt \nonumber \\
			&=\; \frac{2}{\pi} \left[ \text{Si}\left( \frac{N \pi}{2}(\Delta \theta + \mu) \right) - \text{Si}\left( \frac{N \pi}{2}(\Delta \theta - \mu) \right) \right],
		\end{align}
		where $t_{1} = \frac{N\pi}{2}(\Delta \theta+\mu)$, $t_{2} = \frac{N\pi}{2}(\Delta \theta-\mu)$,  and $\text{Si}(a)$ is the sine integral function defined in {\bf Proposition~\ref{Pro:Rolloff}}. Since $|f_{\rm FF}^{(\rm unc)}(\Delta \theta,N)|$ in~\eqref{Exp:Conv_expansion} is an even function w.r.t. $\Delta \theta$, we consider the case where $\Delta \theta \ge 0 $. Using the approximation  $\text{Si}(a)\approx\frac{\pi}{2}$ when $a \ge\pi$ and $\text{Si}(a)\approx -\frac{\pi}{2}$ when $a\le -\pi$, we have $ \text{Si}\left( \frac{N \pi}{2}(\Delta \theta + \mu) \right) \approx \frac{\pi}{2}$ under the condition  $\mu>\frac{2}{N} $ and $\Delta \theta \ge 0 $. Thus,  the above unconstrained beamforming gain $|f_{\rm FF}^{(\rm unc)}(\Delta \theta)|$ can be approximated as 
		\begin{align}
			|f_{\rm FF}^{(\rm unc)}(\Delta \theta,N)| = \frac{2}{\pi}\left[\frac{\pi}{2}-\text{Si}\left( \frac{N \pi}{2}(\Delta \theta - \mu) \right)\right].
		\end{align}
		When $\Delta \theta\le \mu-\frac{2}{N}$ and $\Delta \theta\ge \mu+\frac{2}{N}$, we have $\text{Si}\left( \frac{N \pi}{2}(\Delta \theta - \mu) \right) \approx -\frac{\pi}{2}$	and $\text{Si}\left( \frac{N \pi}{2}(\Delta \theta - \mu) \right) \approx \frac{\pi}{2}$, respectively, thus completing the proof.

		\section{}\label{App:1}
		Let $\varphi_{n}(\theta,\xi) \triangleq u_{n}\theta - \frac{u_{n}^{2}(1-\theta^{2})}{2}\xi$ denote the phase term of $\big[\mathbf{a}_{\rm NF}(\theta,\xi)\big]_{n}$ in~\eqref{Exp:NFCSV}.
		For any point $(\theta,\xi)\in\mathcal{A}_{\rm NF}$, $\varphi_{n}(\theta,\xi)$ can be approximated by its first-order Taylor expansion around $(\theta_{0},\xi_{0})$ as
		\begin{align}\label{Exp:Taylor}
			\varphi_{n}(\theta,\xi) &\approx
			\varphi_{n}(\theta_{0},\xi_{0}) + 
			\frac{\partial \varphi_{n}}{\partial \theta}\Big|_{(\theta_{0},\xi_{0})} \Delta\theta + 
			\frac{\partial \varphi_{n}}{\partial \xi}\Big|_{(\theta_{0},\xi_{0})} \Delta\xi, \nonumber 
		\end{align}
		where $\Delta\theta=\theta-\theta_{0}$ and $\Delta\xi=\xi-\xi_{0}$. By substituting
		$\frac{\partial \varphi_{n}}{\partial \theta} = u_{n}+u_{n}^{2}\theta\xi$
		and
		$\frac{\partial \varphi_{n}}{\partial \xi} = -\frac{u_{n}^{2}(1-\theta^{2})}{2}$
		into~\eqref{Exp:Taylor}, the approximated phase expression is obtained, which completes the proof.

		\section{}\label{App:2}
		To ensure the passband-like beam coverage, the ideal antenna-domain sequence $[\bar{\mathbf{s}}_{\rm NF}]_{n},n\in\mathcal{Z}$ can be obtained via the following inverse transform at the spatial-frequency samples ($\zeta_n^{(\theta)}$ and $\zeta_n^{(\xi)}$) as
		\begin{align}
			 \bar{s}_{{\rm NF},n} &= 
			\iint_{\mathbb{R}^2} {f}_{\rm NF}\,
			e^{\jmath\frac{2\pi}{\lambda}(\zeta_n^{(\theta)}\Delta\theta+\zeta_n^{(\xi)}\Delta\xi)}
			\,d(\Delta\theta)\,d(\Delta\xi) \nonumber \\
			& =  \iint_{-\mu\le \Delta\theta\le \mu \atop -\nu\le \Delta\xi\le \nu}
			{\gamma_{\rm NF}}\,
			e^{\jmath\frac{2\pi}{\lambda}(\zeta_n^{(\theta)}\Delta\theta+\zeta_n^{(\xi)}\Delta\xi)}
			\,d(\Delta\theta)\,d(\Delta\xi)\nonumber \\
			& =  {\gamma_{\rm NF}}
			\Big[\int_{-\mu}^{\mu} e^{\jmath \frac{2\pi}{\lambda}\zeta_{n}^{(\theta)} u } du   \Big] \cdot 
			\Big[\int_{-\nu}^{\nu}e^{\jmath \frac{2\pi}{\lambda}\zeta_{n}^{(\xi)} v  }d v   \Big] \nonumber \\
			& =\!  4 {\gamma_{\rm NF}}
			\Big[\mu\,\mathrm{sinc}  \Big(\frac{2\mu \zeta_n^{(\theta)}}{\lambda}\Big)\Big]
			\Big[\nu\,\mathrm{sinc}\Big(\frac{2\nu\zeta_n^{(\xi)}}{\lambda}\Big)\Big], n\in \mathcal{Z}. \nonumber
		\end{align}
		Let $[\bar{\mathbf{v}}_{\rm NF}]_{n}\triangleq \bar{v}_{{\rm NF},n} = \mathrm{sinc}  \Big(\frac{2\mu \zeta_n^{(\theta)}}{\lambda}\Big)\cdot
		\mathrm{sinc}\Big(\frac{2\nu\zeta_n^{(\xi)}}{\lambda}\Big),n\in\mathcal{Z}$. Similarly, by truncating the ideal infinite sequence $\bar{\mathbf{v}}_{\rm NF}$ with the array mask $\mathbf{m}$ and adding a protective zoom of width $\frac{2}{N}$ in the angle domain to mitigate the roll-off effect, we thus obtain the designed beamforming vector $\mathbf{w}_{\rm NF}$ in~\eqref{Exp:NFBeamvec} based on $ w_{{\rm NF},n} = v_{{\rm NF},n} e^{\jmath \frac{2\pi}{\lambda} \varphi_{n} },\forall n\in \mathcal{N}$, where $\alpha$ is an auxiliary variable introduced to ensure the transmit power constraint.

		\section{}\label{App:beamgain}
		Given any angle deviation $\mu>0$, inverse range deviation $\xi = 0$, and reference point $(\theta_{0},\xi_{0})$, the unconstrained beamforming gain at the point $(\theta_{0},\xi_{0} + \Delta \xi )$ can be rewritten as 
		\begin{align}
			& |f_{\rm NF}^{(\rm unc)}(0,\Delta\xi,N)|  \nonumber \\
			=& \left| {\frac{1}{N}}\sum_{n=1}^{N} \mathrm{sinc}  \Big(\frac{2\mu (u_{n}+u_{n}^{2}\theta_{0} \xi_{0}) }{\lambda}\Big) e^{\jmath \frac{\pi (1-\theta_{0}^{2})\Delta\xi}{\lambda} u_{n}^{2}} 
			\right| \nonumber \\
			=& \left| {\frac{1}{N}}\sum_{n=1}^{N} \mathrm{sinc}  \Big(\kappa (u_{n}+ \varkappa u_{n}^{2}) \Big) e^{\jmath \psi u_{n}^{2}} 
			\right| \triangleq \left|F(\kappa,\varkappa,\psi) \right|,
		\end{align}
		where $\kappa = \frac{2\mu}{\lambda}$, $\varkappa=\theta_{0}\xi_{0}$, and
		$\psi = \frac{\pi (1-\theta_{0}^{2})\Delta\xi}{\lambda}$. The function $F(\kappa,\varkappa,\psi)$ can be approximated as 
		\begin{align}\label{Exp:int1}
			F(\kappa,\varkappa,\psi)&={\frac{1}{N}}\sum_{n=1}^{N} \mathrm{sinc}  \Big(\kappa (u_{n}+ \varkappa u_{n}^{2}) \Big) e^{\jmath \psi u_{n}^{2}}  \nonumber\\
			&\approx \frac{1}{D}\int_{-\frac{D}{2}}^{\frac{D}{2}}\mathrm{sinc} \Big(\kappa x + \kappa \varkappa x^{2} \Big) e^{\jmath \psi x^{2}}\; dx. 
		\end{align}
		Using $\sinc(a)=\frac{1}{2}\int_{-1}^{1} e^{\jmath\pi a t}\;dt$, $F(\kappa,\varkappa,\psi)$ in~\eqref{Exp:int1} can be re-expressed as
		\begin{align}\label{Exp:int2}
			&\quad F(\kappa,\varkappa,\psi) = \frac{1}{D}\int_{-\frac{D}{2}}^{\frac{D}{2}} \Big[\frac{1}{2}\int_{-1}^{1} e^{\jmath\pi (\kappa x + \kappa \varkappa x^{2}) t}\;dt\Big] e^{\jmath \psi x^{2}}\; dx\nonumber\\
			&= \frac{1}{2D}\int_{-1}^{1} \underbrace{ \left[ \int_{-\frac{D}{2}}^{\frac{D}{2}} e^{\jmath \left[ (\psi + \pi \kappa \varkappa t)x^2 + (\pi \kappa t)x \right]} \; dx \right] }_{\mathcal{I}(t)} \; dt.
		\end{align}
		Let $\mathcal{Q}(t) = \psi + \pi \kappa \varkappa t$ and $\mathcal{J}(t) = \pi \kappa t$. The inner integral in~\eqref{Exp:int2} can be recognized as a generalized Fresnel integral, i.e.,
		\begin{align}\label{Exp:I_t_pre}
			\mathcal{I}(t) &= e^{-\jmath \frac{\mathcal{J}^2(t)}{4\mathcal{Q}(t)}} \int_{-\frac{D}{2}}^{\frac{D}{2}} e^{\jmath \mathcal{Q}(t) \left(x + \frac{\mathcal{J}(t)}{2\mathcal{Q}(t)}\right)^2} \; dx.
		\end{align}
		{\bf Case 1}~$(\mathcal{Q}(t) > 0)$:
		Let $\tau = \sqrt{\frac{2\mathcal{Q}(t)}{\pi}} \left(x + \frac{\mathcal{J}(t)}{2\mathcal{Q}(t)}\right)$. Then, The integral \eqref{Exp:I_t_pre} is rewritten as 
		\begin{align}
			\mathcal{I}(t) &= \sqrt{\frac{\pi}{2\mathcal{Q}(t)}} e^{-\jmath \frac{\mathcal{J}^2(t)}{4\mathcal{Q}(t)}} \int_{z_1(t)}^{z_2(t)} e^{\jmath \frac{\pi}{2} \tau^2} \; d\tau \nonumber \\
			&= \sqrt{\frac{\pi}{2\mathcal{Q}(t)}} e^{-\jmath \frac{\mathcal{J}^2(t)}{4\mathcal{Q}(t)}} \Big[ \mathcal{G}(z_2(t)) - \mathcal{G}(z_1(t)) \Big],
		\end{align}
		where $\mathcal{G}(z) = \mathcal{C}(z) + \jmath\mathcal{S}(z)$, $z_1(t) = \sqrt{\frac{2\mathcal{Q}(t)}{\pi}} \left( - \frac{D}{2} + \frac{\mathcal{J}(t)}{2\mathcal{Q}(t)} \right)$, and $ z_2(t) = \sqrt{\frac{2\mathcal{Q}(t)}{\pi}} \left( \frac{D}{2} + \frac{\mathcal{J}(t)}{2\mathcal{Q}(t)} \right)$. Herein, $\mathcal{C}(z) = \int_{0}^{z} \cos\left(\frac{\pi}{2}t^2\right) dt$ and $\mathcal{S}(z) = \int_{0}^{z} \sin\left(\frac{\pi}{2}t^2\right) dt$ are Fresnel
		functions.
		{\bf Case 2}~$(\mathcal{Q}(t) < 0)$:
		Let $\tau = \sqrt{\frac{-2\mathcal{Q}(t)}{\pi}} \left(x + \frac{\mathcal{J}(t)}{2\mathcal{Q}(t)}\right)$. Then, the integral \eqref{Exp:I_t_pre} is rewritten as 
		\begin{align}
			\mathcal{I}(t) &= \sqrt{\frac{\pi}{-2\mathcal{Q}(t)}} e^{-\jmath \frac{\mathcal{J}^2(t)}{4\mathcal{Q}(t)}} \int_{z'_1(t)}^{z'_2(t)} e^{-\jmath \frac{\pi}{2} \tau^2} \; d\tau \nonumber \\
			&= \sqrt{\frac{\pi}{-2\mathcal{Q}(t)}} e^{-\jmath \frac{\mathcal{J}^2(t)}{4\mathcal{Q}(t)}} \Big[ \mathcal{G}^*(z'_2(t)) - \mathcal{G}^*(z'_1(t)) \Big],
		\end{align}
		where $z'_1(t) = \sqrt{\frac{-2\mathcal{Q}(t)}{\pi}} \left( - \frac{D}{2} + \frac{\mathcal{J}(t)}{2\mathcal{Q}(t)} \right)$ and 
		$z'_2(t) = \sqrt{\frac{-2\mathcal{Q}(t)}{\pi}} \left( \frac{D}{2} + \frac{\mathcal{J}(t)}{2\mathcal{Q}(t)} \right)$.
		Combining the above two cases, and letting $z_1(t) = \sqrt{\frac{2|\mathcal{Q}(t)|}{\pi}} \left( - \frac{D}{2} + \frac{\mathcal{J}(t)}{2\mathcal{Q}(t)} \right)$ as well as 
		$z_2(t) = \sqrt{\frac{2|\mathcal{Q}(t)|}{\pi}} \left( \frac{D}{2} + \frac{\mathcal{J}(t)}{2\mathcal{Q}(t)} \right)$, the integral \eqref{Exp:I_t_pre} is given by 
		\begin{align}
			\mathcal{I}(t) &= \sqrt{\frac{\pi}{2|\mathcal{Q}(t)|}} e^{-\jmath \frac{\mathcal{J}^2(t)}{4\mathcal{Q}(t)}} \times \nonumber \\
			&\begin{cases}
				\big[ \mathcal{G}(z_2(t)) - \mathcal{G}(z_1(t)) \big], & \mathcal{Q}(t) > 0 \\
				\big[ \mathcal{G}^*(z_2(t)) - \mathcal{G}^*(z_1(t)) \big], & \mathcal{Q}(t) < 0
			\end{cases}.
		\end{align}
		As such, by exploiting the symmetry of the outer integral, $F(\kappa,\varkappa,\psi)$ is given by $	F(\kappa,\varkappa,\psi) \approx \frac{1}{2D} \int_{0}^{1} [ \mathcal{I}(t) + \mathcal{I}(-t) ] \; dt$,
		thus completing the proof.
		
	\end{appendices}

	\bibliographystyle{IEEEtran}
	\bibliography{Ref_title.bib}
	
\end{document}